\documentclass[10pt]{article}
\usepackage[margin=1in]{geometry}

\usepackage{soul}
\usepackage{color}
\usepackage{caption}
\usepackage{subcaption}
\usepackage{subfiles}
\usepackage{subcaption}
\usepackage{graphicx}
\usepackage{comment}
\usepackage{multirow}
\usepackage{verbatim}
\usepackage{listings}
\usepackage[toc]{appendix}
\usepackage{pifont,xspace}
\usepackage{url}
\usepackage{hyperref}
\usepackage{authblk}

\usepackage[htt]{hyphenat}

\newcommand{\oneCircled}{\ding{182}\xspace}
\newcommand{\twoCircled}{\ding{183}\xspace}
\newcommand{\threeCircled}{\ding{184}\xspace}
\newcommand{\fourCircled}{\ding{185}\xspace}
\newcommand{\fiveCircled}{\ding{186}\xspace}
\newcommand{\sixCircled}{\ding{187}\xspace}

\definecolor{lightGreen}{rgb}{0.40,0.73,0.21}
\newcommand*{\lgText}[1]{\textcolor{lightGreen}{#1}}

\title{Traffic Modeling with SUMO: a Tutorial}

\author[1,2]{Davide Andrea Guastella\thanks{davide.guastella@lis-lab.fr}}
\author[2]{Eladio Montero-Porras\thanks{eladio.montero.porras@ulb.be}}
\author[2]{Alejandro Morales-Hern\'{a}ndez\thanks{alejandro.morales.hernandez@ulb.be}}
\author[2]{Gianluca Bontempi\thanks{gianluca.bontempi@ulb.be}}

\affil[1]{Aix Marseille University, CNRS, LIS, Marseille, France}
\affil[2]{Machine Learning Group, Universit\'{e} Libre de Bruxelles}

\date{}

\newcommand{\dollar}{\mbox{\textdollar}}

\begin{document}
\maketitle

\begin{abstract}
This paper presents a step-by-step guide to generating and simulating a traffic scenario using the open-source simulation tool SUMO. It introduces the common pipeline used to generate a synthetic traffic model for SUMO, how to import existing traffic data into a model to achieve accuracy in traffic simulation (that is, producing a traffic model which dynamics is similar to the real one). It also describes how SUMO outputs information from simulation that can be used for data analysis purposes.
\end{abstract}

\section{Introduction}

The concept of Smart Transportation has become more predominant over the past decade, encompassing innovative methods for the reduction of traffic congestion, traffic accidents, and air pollution, all of which engender excessive costs to society and impact the general well-being of citizens. This is necessary because of the increasing number of vehicles in urban environments. Although the awareness of city governments about sustainable mobility, such as investing in the design and development of mass transportation systems to reduce CO$_2$ emissions, still the high number of vehicles makes it necessary to analyze and implement policies for urban infrastructure management to optimally convey traffic and avoid congestion and the consequent air pollution. On the one hand, the impact of the control strategies on road infrastructures is not observable until they are deployed in the real world~\cite {KUSIC2023101858}. On the other hand, testing control strategies in real-life settings are expensive, risky, and often unfeasible~\cite{argota_sanchez_vaquerizo_2021}. In this context, urban traffic simulation models have become an indispensable asset, providing an \textit{in-silico} environment where it is possible to design and assess alternative control strategies before the deployment. In this context, urban traffic simulation models have become an indispensable asset: these tools provide a lens through which it is possible to analyse control strategies \textit{in silico}. They rely on computational models to test these strategies before deploying them in the real world.





Some of the advantages of traffic simulators are detailed as follows~\cite{dorokhin_traffic_2020}: 

\begin{itemize}
	\item Using a simulation model, traffic management experts can assess decision-making tasks to reduce traffic congestion of certain sections of roads.
	\item A simulation model relies on an accurate topography of the city, consisting of buildings, roads, intersections, bridges. Using appropriate modeling software, it is possible to improve the geometric design of the road and see how these changes will affect the typical traffic flow.
	\item A simulation model allows evaluating the time and cost of the trip of vehicles. This is important when it is necessary to determine the economic assessment of a road infrastructure change. A specialist planning the transport work can conduct a comparative assessment of diverse options for traffic routes without significant material and time costs.
\end{itemize}

One of the challenges in traffic simulation lies in creating an accurate model of urban traffic: this requires several types of data, which are not always publicly available. The required data include socio-economic indicators as well as historical information about traffic flow. Having such information is crucial for defining accurate synthetic traffic models.

The goal of this document is to introduce traffic scenario modeling with the open-source simulation tool SUMO.

SUMO (Simulation of Urban MObility) is an open-source traffic simulation software designed for modeling and analyzing transportation systems. Developed by the German Aerospace Center (DLR), SUMO allows researchers, urban planners, and engineers to simulate realistic traffic scenarios, including private and public transport, pedestrian movement, and traffic light systems. It provides a microscopic simulation approach, meaning it models individual vehicle behavior based on car-following and lane-changing models. SUMO supports various input data formats, including real-world road networks imported from OpenStreetMap (OSM) and demand data from different sources, making it a versatile tool for traffic research and analysis.

One of SUMO's key advantages is its flexibility in integrating with external applications, enabling users to test traffic control algorithms, intelligent transportation systems (ITS), and autonomous vehicle coordination strategies. It provides detailed outputs on traffic dynamics, such as travel times, emissions, and congestion patterns for evaluating transportation policies and infrastructure planning. 

The rest of the document is organized as follows: Section~\ref{sec:intro} introduces the common pipeline used to generate a synthetic traffic model for SUMO. Section~\ref{sec:output} describes how SUMO outputs information from simulations, which can be used for traffic data analysis. Section~\ref{sec:tools} introduces the main tools that enable the automatic generation of synthetic traffic models for SUMO.

\section{Creating a Synthetic Traffic Scenario}\label{sec:intro}

\begin{figure}[!ht]
    \centering
    \includegraphics[width=.85\columnwidth]{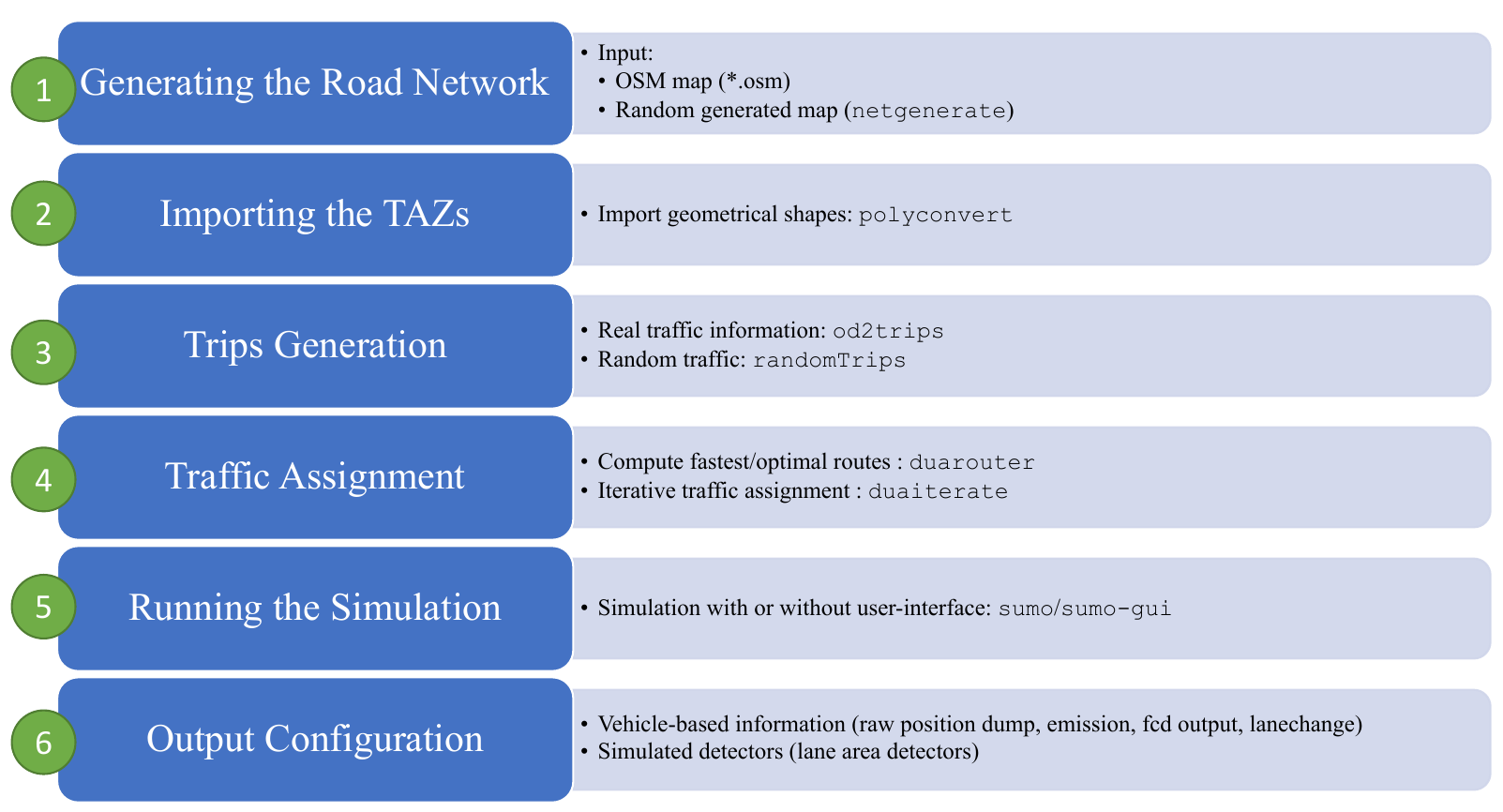}
    \caption{Main steps required to generate and simulate a traffic scenario with SUMO.}
    \label{fig:random_scenario_steps}
\end{figure}

SUMO enables generating random road networks or converting OpenStreetMap (OSM) extracts to a specific format that can be used in SUMO, so to have a real-work representation of a road network. We will also focus on modeling synthetic traffic flow, which can be generated from either existing information about real traffic or generated randomly. Figure~\ref{fig:random_scenario_steps} shows the overall steps required to generate a synthetic traffic scenario using SUMO. Each step will be discussed in the following sections.

Following, we briefly describe the steps required to generate and simulate a traffic scenario:

\begin{itemize}
	\item[\lgText{\oneCircled}] \textbf{Importing the network}: this step produces a road network model that can be used in SUMO. The road network can be either generated randomly or converted from OpenStreetMap;
	\item[\lgText{\twoCircled}] \textbf{Importing the Traffic Analysis Zones (TAZs)}: import traffic analysis zones definitions. TAZs are polygons that delimits an urban environment according to some socio-economical indicators (such as average income, education level, traffic pressure, or simply administrative boundaries). TAZs are typically defined by governmental organizations. Dividing the environment into TAZs is useful to model the traffic flow from one local area of a city to another one;	
	\item[\lgText{\threeCircled}] \textbf{Trips Generation}: assign an origin and a destination to each vehicle that should be inserted into the simulation. This can be done using either real or synthetic traffic data;
	\item[\lgText{\fourCircled}] \textbf{Traffic Assignment}: model the routes (herein a route is the complete path for going from origin to destination) for each vehicle in the simulation;
	\item[\lgText{\fiveCircled}] \textbf{Simulation}: simulate traffic  using the road network and the defined traffic model;
	\item[\lgText{\sixCircled}] \textbf{Output Configuration}: output information generated from the simulation that can be used for traffic analysis purposes;
\end{itemize}

\subsection{Road Network Generation (step~\lgText{\oneCircled})}

This section introduces the tools available in SUMO to create a random road network and to import a road network from OpenStreetMap.

\subsubsection{Random Road Networks}

The command \verb+netgenerate+\footnote{\href {https://sumo.dlr.de/docs/netgenerate.html}{https://sumo.dlr.de/docs/netgenerate.html}} allows generating three types of abstract road networks: grid (using \verb+--grid+ parameter), spider (using \verb+--spider+ parameter) and random (using \verb+--random parameter+). The use of randomly generated road networks is pertinent to validate data analysis techniques in traffic domain; the same technique that rely on traffic data can be evaluated on several simulated scenarios, each one having a specific topology, type of junctions, or traffic light options.

\begin{figure}[!ht]
    \centering
    \includegraphics[width=.6\columnwidth]{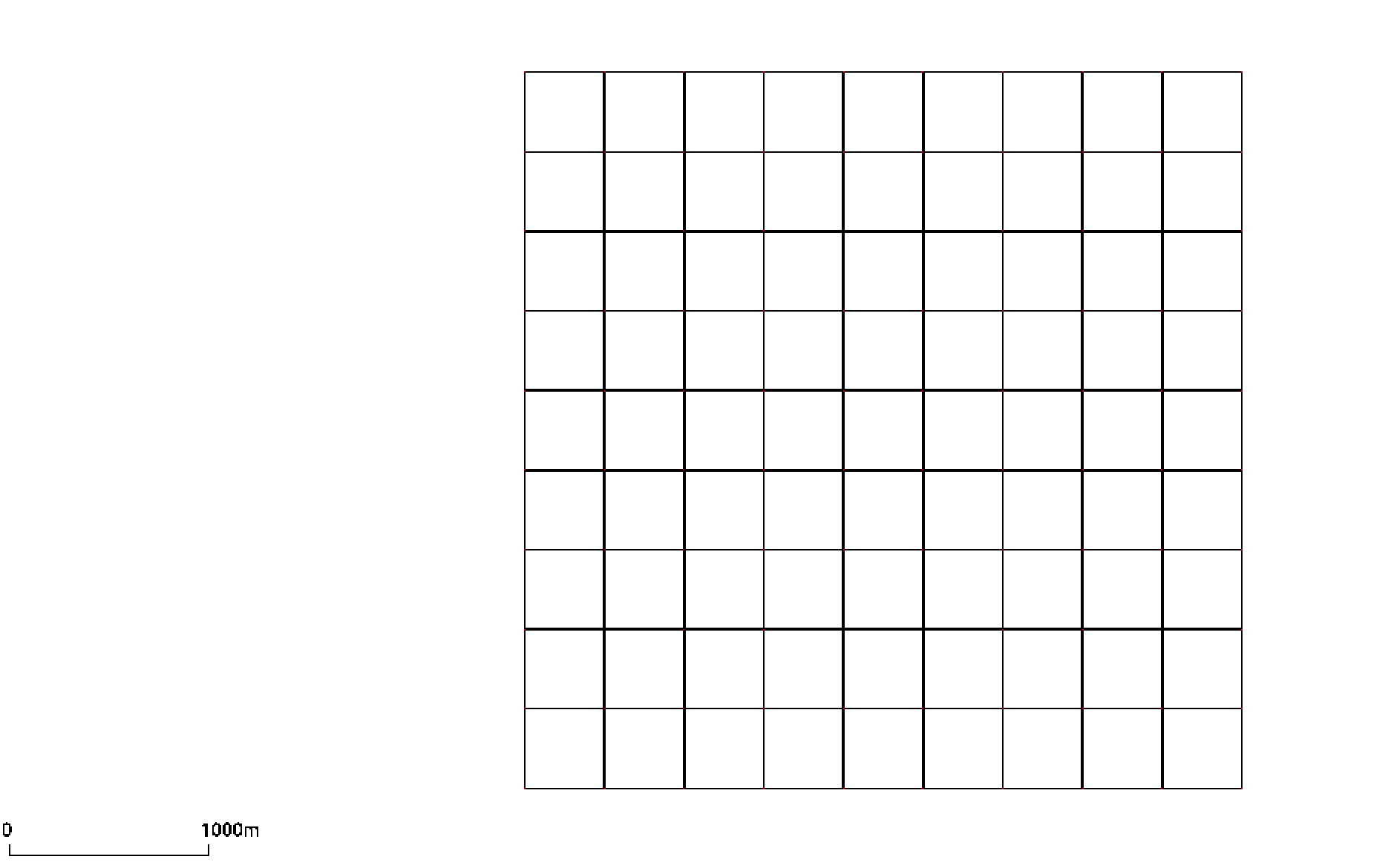}
    \caption{Random grid road network topology generated by \texttt{netgenerate}.}
    \label{fig:random_grid}
\end{figure}

The following command creates a random grid-like road network topology:

\begin{lstlisting}
> netgenerate --grid --grid.number=10 --grid.length=400 
	--output-file=MySUMOFile.net.xml	
\end{lstlisting}

\noindent which produces the output shown in Figure~\ref{fig:random_grid}.

\begin{figure}[!ht]
    \centering
    \includegraphics[width=.6\columnwidth]{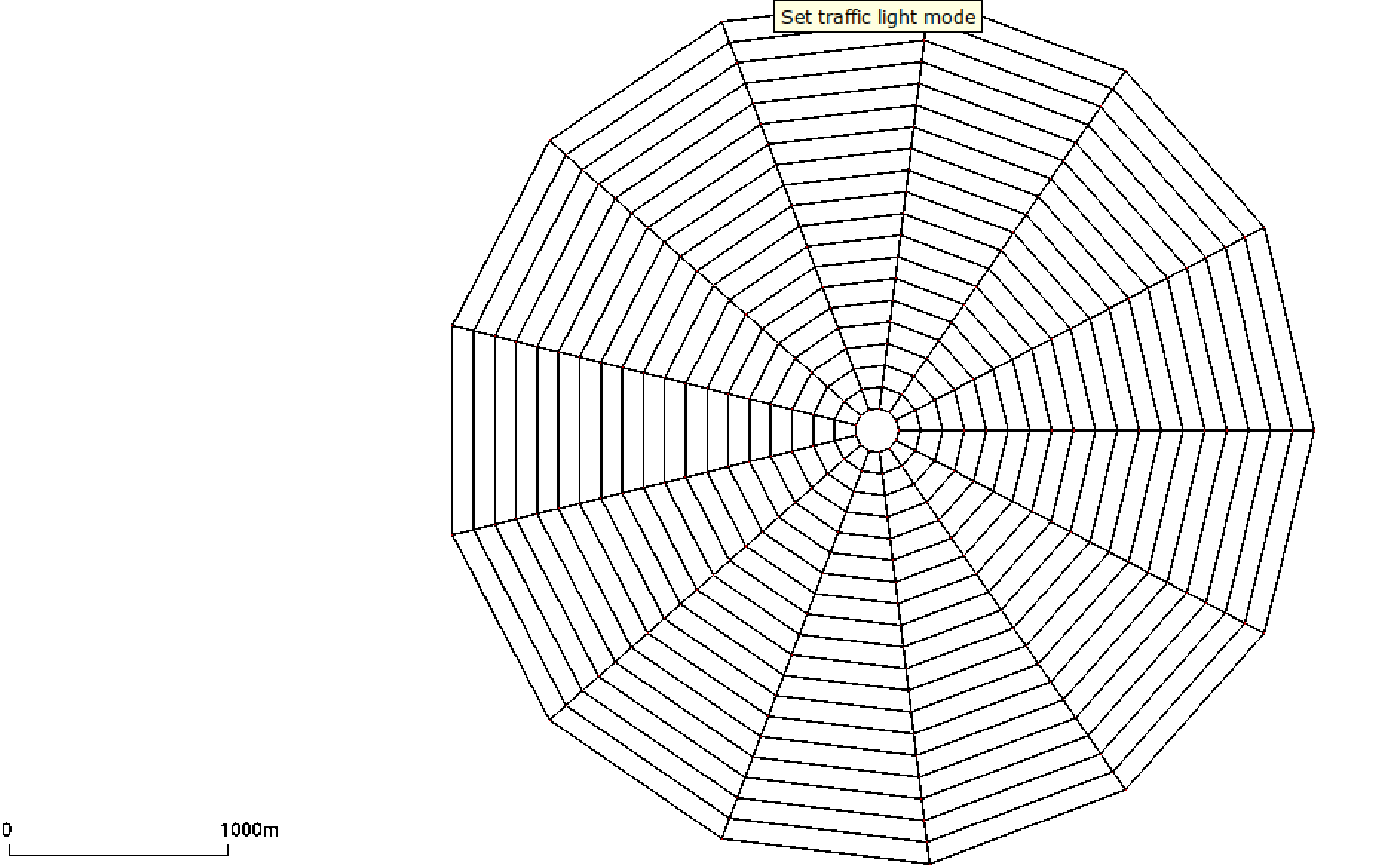}
    \caption{Random spider-like road network topology generated by \texttt{netgenerate}.}
    \label{fig:random_spider}
\end{figure}

The following command creates a random spider-like road network topology:

\begin{lstlisting}
> netgenerate --spider --spider-omit-center --output-file=MySUMOFile.net.xml
\end{lstlisting}

\noindent which produces the output shown in Figure~\ref{fig:random_spider}.

\begin{figure}[!ht]
    \centering
    \includegraphics[width=.6\columnwidth]{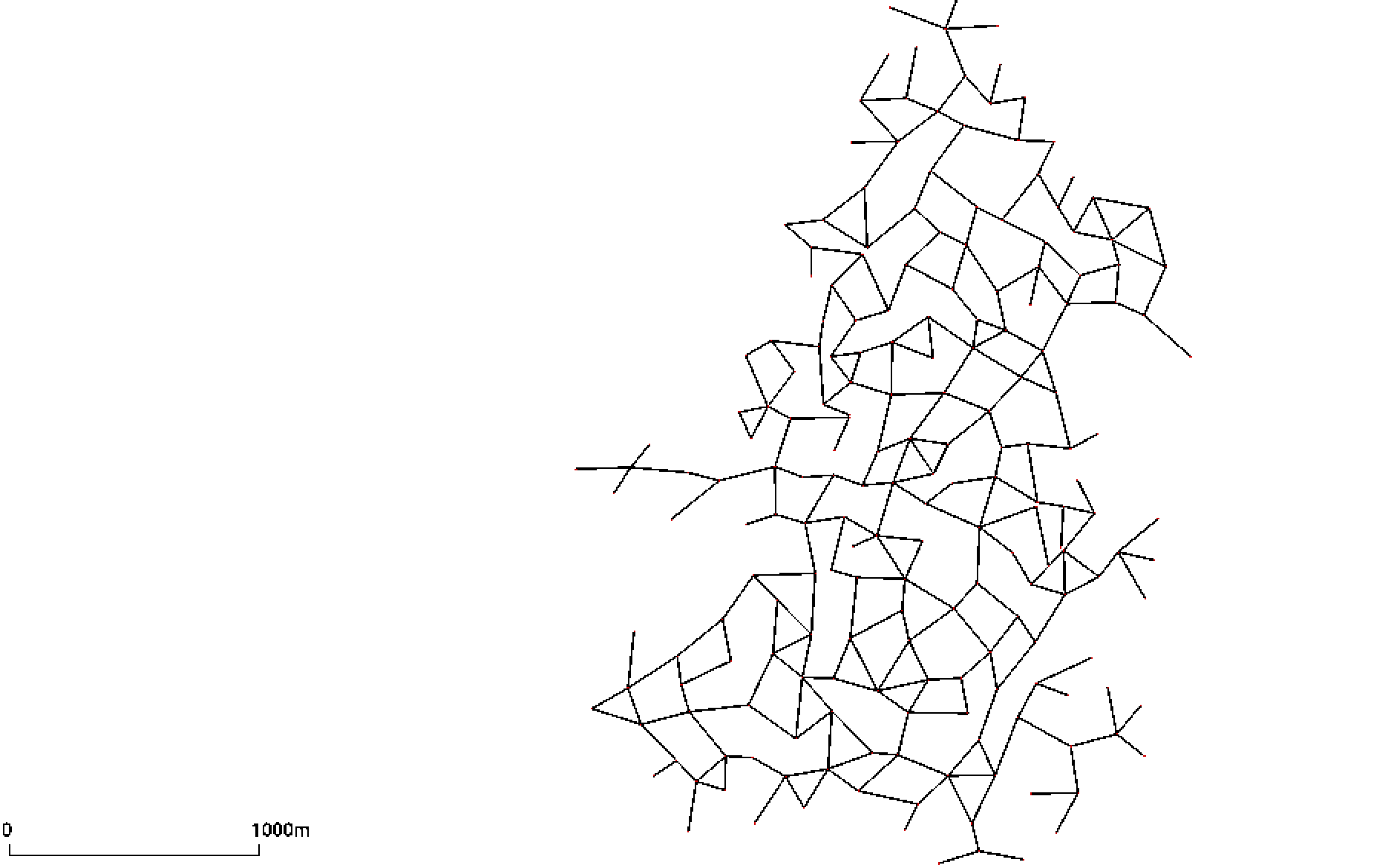}
    \caption{Random road network topology generated by \texttt{netgenerate}.}
    \label{fig:random_random}
\end{figure}

The following command creates a random road network topology:

\begin{lstlisting}
> netgenerate --rand -o MySUMOFile.net.xml --rand.iterations=200
\end{lstlisting}

\noindent which produces the output shown in Figure~\ref{fig:random_random}.

\begin{figure}[!ht]
    \centering
    \includegraphics[width=.6\columnwidth]{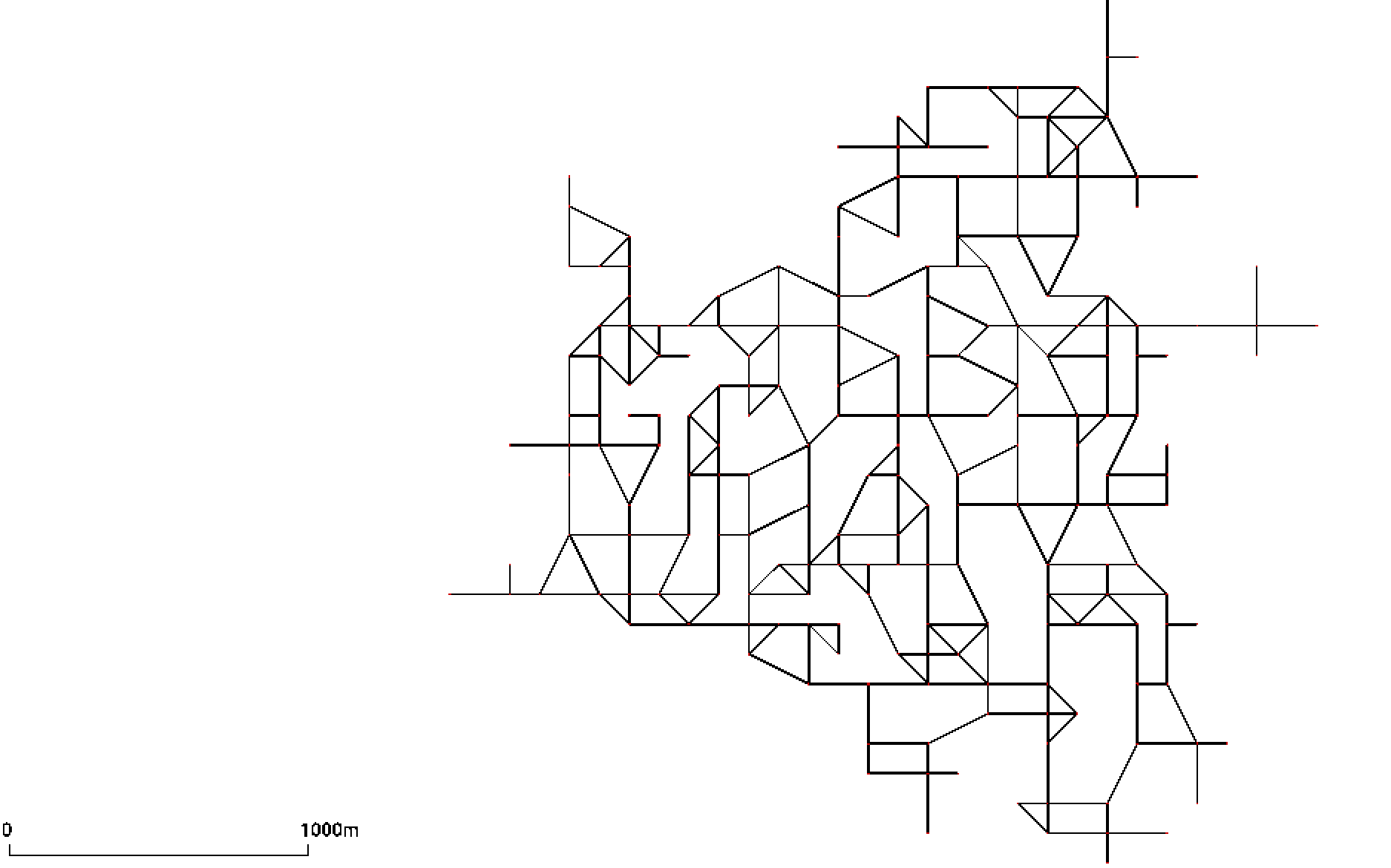}
    \caption{}
    \label{fig:random_randgrid}
\end{figure}

Additionally, by setting the option \verb+--rand.grid+, additional grid structure is enforce during random network generation, which produces the output shown in Figure~\ref{fig:random_randgrid}.

\subsubsection{Extracting a Road Network Topology from OpenStreetMap}

In SUMO, a model of a real road network can be defined manually. To do this, an XML file must be defined, containing elements such as the roads definition, intersections, and traffic lights. However, manually defining the road network topology for a real urban area is impractical due to the significant amount of time required. To tackle this issue, SUMO provides a tool to convert road network from OpenStreetMap (OSM) files.

\begin{figure}[!ht]
    \centering
    \includegraphics[width=.85\columnwidth]{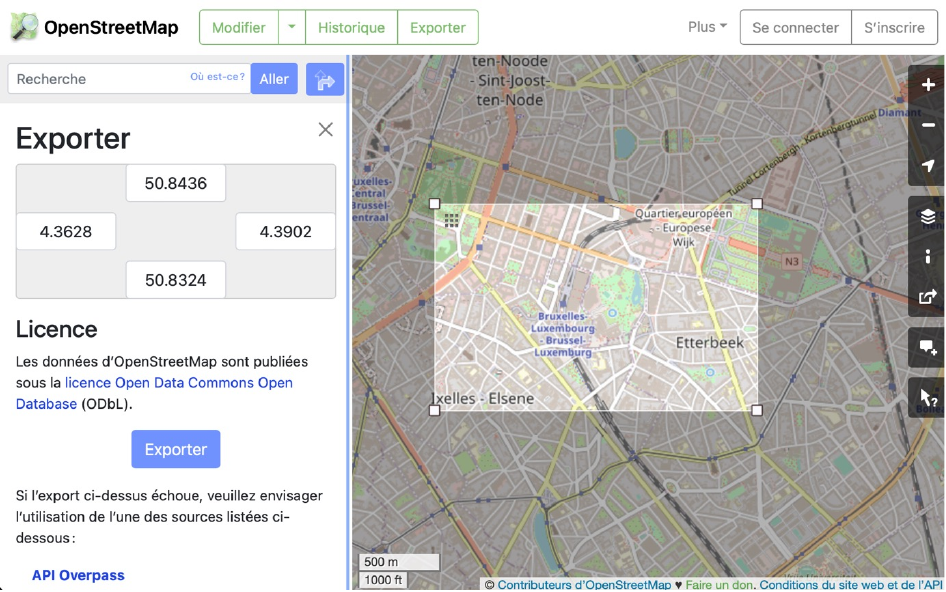}
    \caption{OSM allows exporting a selected area into an XML format.}
    \label{fig:osm_map_extraction}
\end{figure}

The first step to import a road network from OSM in SUMO is to select the region to be modeled, as shown in Figure~\ref{fig:osm_map_extraction}.

The next step is to export and download the map. If the ``Export'' button is selected and the selected area is too big, then OSM will not export the selected area because the number of nodes within the area is above the limit (50000 is the maximum allowed nodes in a selected region for exporting). By selecting the ``overpass API'' link it is possible to overcome this problem. The output is an XML file containing the definition of all the features in the selected region. In OSM, a feature is any physical element (natural or human-made) in the landscape. Some examples of features are buildings, roads, vegetation, land use, railways, and waterways.

In some cases, it is required to model a specific part of the real environment. For this, exporting the OSM file from the website as discussed previously is not a pertinent solution, as this can include parts of the real environment that must not be modeled. Suppose we must model the road network for a part of the Ixelles municipality (Belgium, Figure~\ref{fig:ixelles_highlight}):

\begin{figure}
    \centering
    \includegraphics[width=0.5\linewidth]{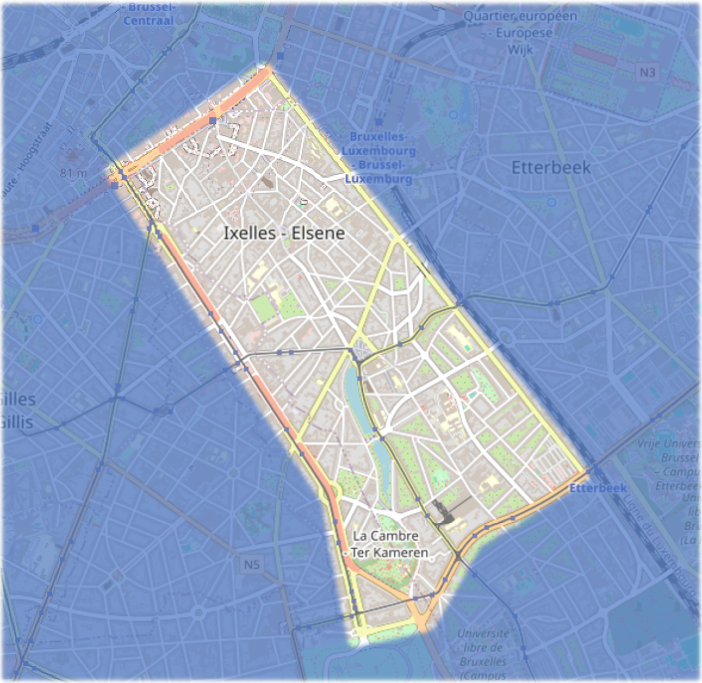}
    \caption{Part of the Ixelles municipality, Belgium}
    \label{fig:ixelles_highlight}
\end{figure}

\begin{figure}
    \centering
    \includegraphics[width=0.8\linewidth]{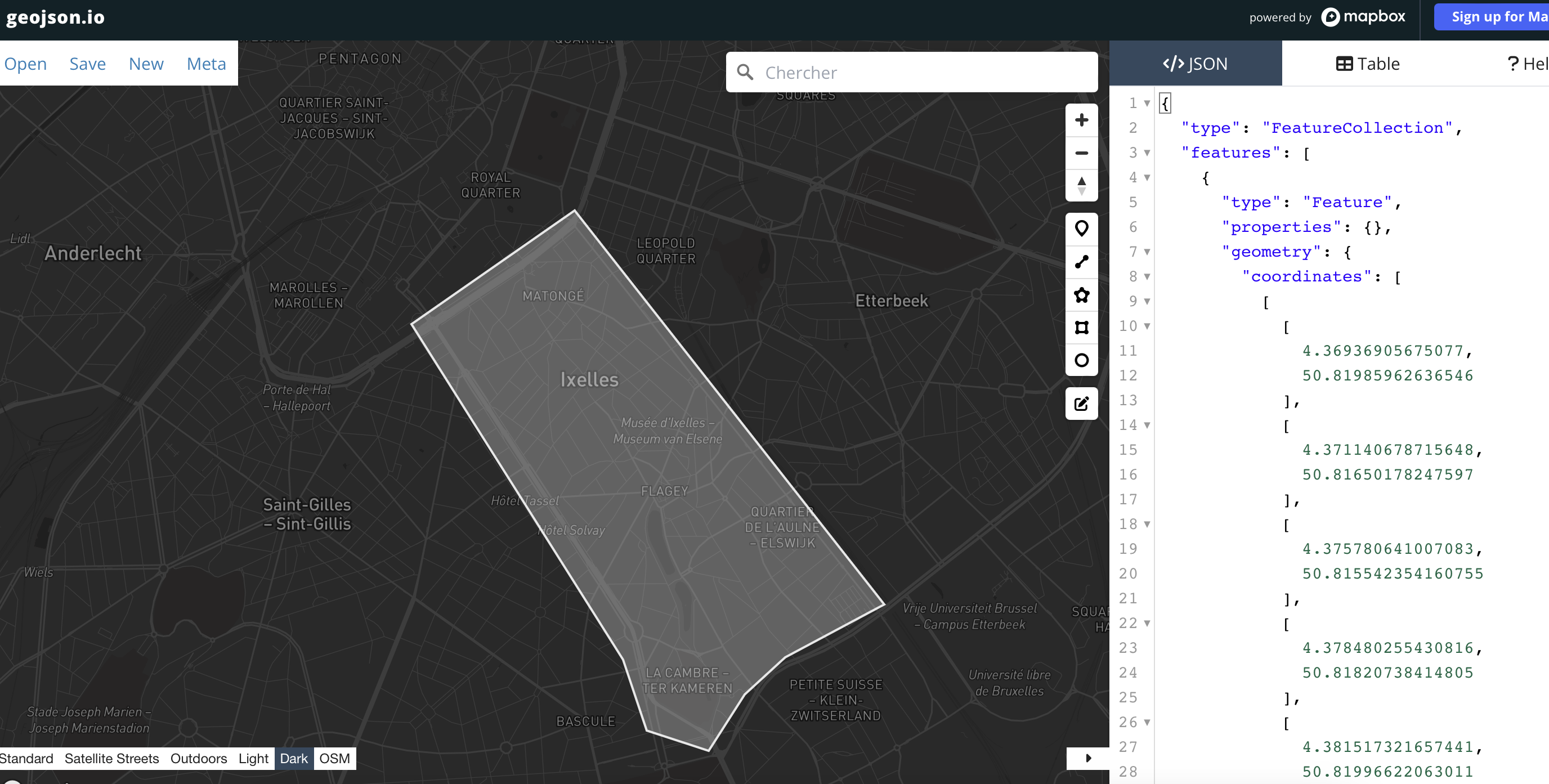}
    \caption{Modelling a specific part of the real environment using a geoJSON online tool.}
    \label{fig:ixelles_geojson}
\end{figure}

By using the OSM tool, it is not possible to obtain a network containing only the roads included in the part of environment in Figure~\ref{fig:ixelles_highlight}. To extract the road network for a specific part of the environment, first define the polygon in geoJSON delimiting the road network to model. Figure~\ref{fig:ixelles_geojson} shows a polygon enclosing a part of the Ixelles municipality. Any online geoJSON modeling tool can be used for this purpose. Herein we use \url{geojson.io}.

After generating the geoJSON containing the polygon that delimitates the road network to model, the following steps must be executed to obtain a road network for SUMO:

\begin{itemize}
    \item Download the \texttt{*.pbf} file of the country that includes the road network part to model (for Belgium the file is available at \url{https://download.geofabrik.de/europe/belgium.html})
    \item Install \texttt{osmium} and \texttt{osmfilter}. In Mac Os X, can be installed through \texttt{brew}\footnote{\href {https://brew.sh}{https://brew.sh}}.
    \item{Extract the \texttt{*.pbf} containing only the elements included in the defined polygon:

    \noindent\begin{verbatim}
osmium extract -p my_poly.geojson my_country.osm.pbf  
    -o filtered.pbf --set-bounds --overwrite
    \end{verbatim}

    \noindent where \texttt{my\_poly.geojson} is the geoJSON file, \texttt{my\_country.osm.pbf} is the file of the modelled country, \texttt{filtered.pbf} is the filtered country \texttt{*.pbf} file referring to only the part included in the defined polygon.}
    \item{Convert \texttt{*.pbf} to \texttt{*.osm}:

    \begin{verbatim}
osmium cat filtered.pbf -o road_net.osm
    \end{verbatim}
    }
    \item{
    Filter only the road network. All unnecessary elements such as buildings or walkways are discarded:

    \begin{verbatim}
osmfilter road_net.osm --keep="highway=* building=*" -o=road_net_filtered.osm
    \end{verbatim}
    }
    \item{Finally, the output file \texttt{road\_net\_filtered.osm} can be converted to SUMO format using the \texttt{netconvert} tool:
    
    \begin{verbatim}
netconvert --osm road_net_filtered.osm  -o net.xml --ramps.guess 
    --junctions.join --remove-edges.isolated --output.street-names 
    --output.original-names
    \end{verbatim}
    }
\end{itemize}

All the features in the OSM XML file are represented as ``\texttt{node}'' tags. Each node has an identifier, a position (in latitude/longitude), and other useful information that allows identifying the type of node, such as the type of road, building (apartments, offices, theaters, etc.), or bike lane.

Before converting the OSM file into a format compatible with SUMO, it may be useful for decision-making purposes to modify certain properties of the road network, such as the direction of the roads. Appendix~\ref{app:reverse_road} includes the Python code to reverse the direction of one or more roads in an OSM-format road network.

To use an OSM map in SUMO, it is necessary to convert the XML file containing the definition of the extracted urban area to a specific format for use with the simulator. SUMO comes with a tool named \verb+netconvert+\footnote{\href {https://sumo.dlr.de/docs/netconvert.html}{https://sumo.dlr.de/docs/netconvert.html}}, that allows importing road networks from different sources such as:

\begin{itemize}
\item ``SUMO plain'' XML descriptions (*.edg.xml, *.nod.xml, *.con.xml, *.tll.xml)
\item OpenStreetMap (*.osm.xml/*.osm), including shapes (see OpenStreetMap import)
\item VISUM, including shapes and demands
\item Vissim, including demands
\item OpenDRIVE
\item MATsim
\item SUMO (*.net.xml)
\item Shapefiles (.shp, .shx, .dbf), e.g. ArcView and newer Tiger networks
\item Robocup Rescue League, including shapes
\item a DLR internal variant of Navteq's GDF (Elmar format)
\end{itemize}

In its most simple usage, netconvert takes in input only the XML file obtained from OSM (parameter \verb+--osm+), and output a road network in XML file (parameter \verb+-o+):

\begin{lstlisting}
> netconvert --osm my_osm_net.xml -o my_sumo_net.net.xml
\end{lstlisting}

We suggest using \texttt{netconvert} with the following options:

\begin{itemize}
    \item \texttt{--ramps.guess}: Enable ramp-guessing.
    \item \texttt{--junctions.join}: Joins junctions that are close to each other.
    \item \texttt{--tls.guess-signals}: Interprets tls nodes surrounding an intersection as signal positions for a larger TLS.
    \item \texttt{--tls.discard-simple}: Does not instantiate traffic lights at geometry-like nodes loaded from other formats than plain-XML.
    \item \texttt{--tls.join}: Try to cluster tls-controlled nodes.
    \item \texttt{--tls.default-type actuated}: Use traffic light programs that adapt to demand dynamically.
    \item \texttt{-t \$SUMO\_HOME/data/typemap/osmNetconvert.typ.xml}: use standard types for converting to SUMO format.
    \item \texttt{--remove-edges.by-vclass rail\_slow,rail\_fast,bicycle,pedestrian}: Remove edges where trains, bikes and pedestrian are allowed. This keeps only the edges for vehicular traffic.
    \item \texttt{--remove-edges.isolated}: remove isolated edges in the road network.
    \item \texttt{--output.street-names}: Street names will be included in the output.
    \item \texttt{--output.original-names}: Keep the original names of the streets in OSM.
\end{itemize}

\subsection{Traffic Assignment Zones (TAZs, step~\lgText{\twoCircled})} \label{sec:taz}

Traffic Analysis Zones (TAZs) are spatial units used in transportation modeling and traffic studies to represent areas with similar travel behavior. A TAZ typically models a neighborhood or district, and serves as a fundamental unit for estimating travel demand. Each zone can be associated with socio-economic and demographic data, such as population, employment, and land use. This information influences the number and type of trips generated within the area. TAZs are commonly used in trip-based models, such as the traditional four-step travel demand model (trip generation, trip distribution, mode choice, and route assignment).

\begin{figure}[!ht]
    \centering
    \includegraphics[width=\textwidth]{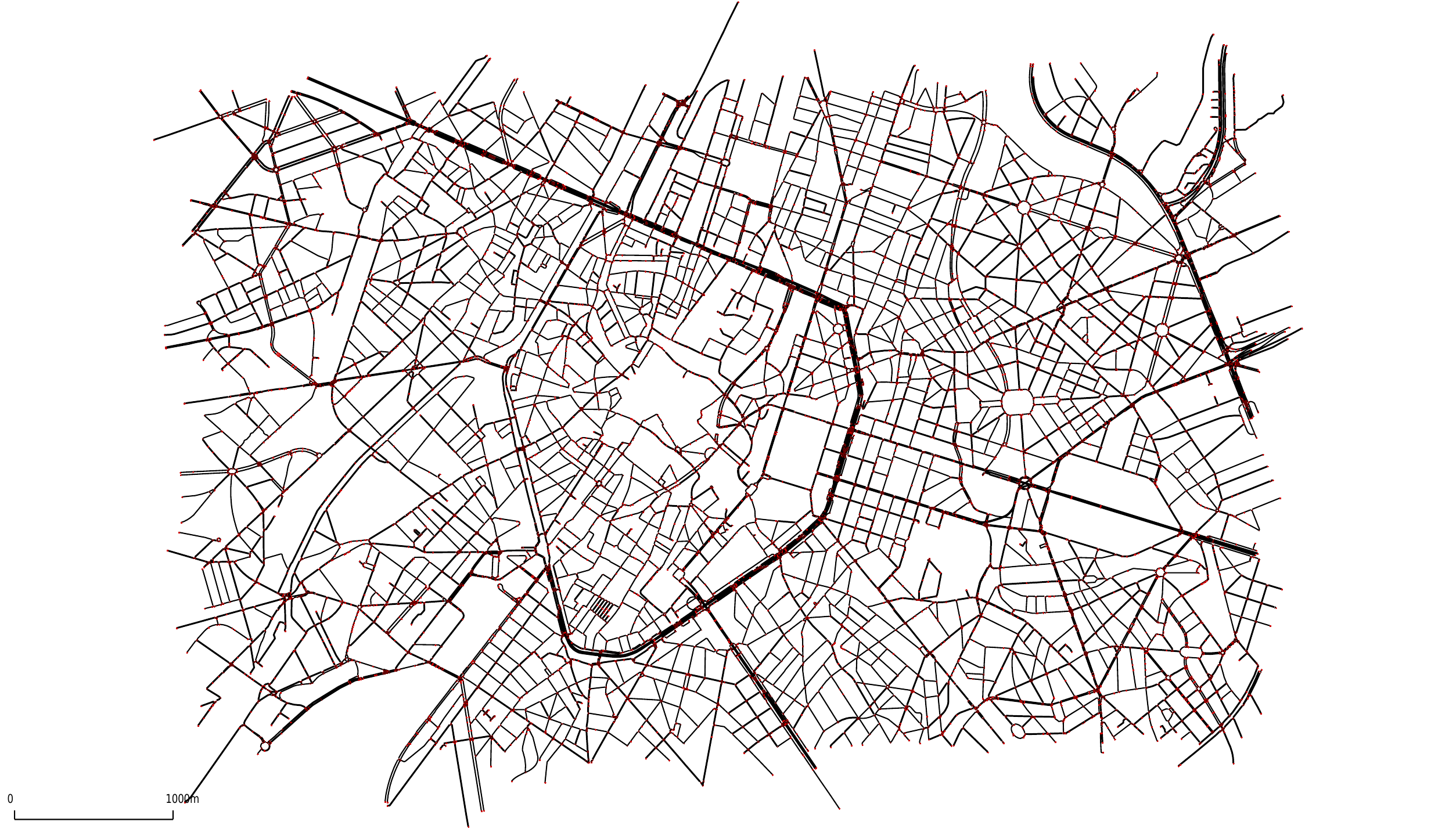}
    \caption{Road network of the city of Brussels.}
    \label{fig:bxl}
\end{figure}

Figure~\ref{fig:bxl} shows the inner part of the route network of the city of Brussels, extracted from OSM and imported in SUMO using the \verb+netconvert+ tool. The area covers approximately an area of 24.000m$^2$. For sake of simplicity, we model only the road network, therefore excluding railways and other buildings.

We now evaluate the TAZs for the modeled area using the spatial boundaries of Brussels Capital Region's neighborhoods. This data is freely available from the computer center for the Brussels Region (Centre d'Informatique pour la R\'{e}gion Bruxelloise, CIRB)\footnote{\href {https://data.metabolismofcities.org/library/33895/}{https://data.metabolismofcities.org/library/33895/}}. The data is provided as shapefile, a geo-spatial vector data format commonly used for Geographic Information System (GIS) software.

To use the shapefile in SUMO as TAZs, this must be converted in a proper format using the \verb+polyconvert+\footnote{\href {https://sumo.dlr.de/docs/polyconvert.html}{https://sumo.dlr.de/docs/polyconvert.html}} tool. This tool can be used to generate additional files for SUMO containing information about all the polygons (e.g., buildings, grounds, etc.).

The following command extracts the polygons from a shapefile and convert them in a format compatible with SUMO. The value ``UrbAdm\_Monitoring\_District'' is the prefix of the shapefile from CIRB, containing the spatial boundaries of Brussels neighborhoods.

\begin{lstlisting}
> polyconvert --shapefile-prefix UrbAdm_Monitoring_District 
	--shapefile.guess-projection true --shapefile.traditional-axis-mapping true 
	--shapefile.id-column ID -n ../bxl.net.xml -o blx.poly.xml
\end{lstlisting}

\noindent where \verb+UrbAdm_Monitoring_District+ is the prefix of the shapefile. The parameter \verb+--shapefile.guess-projection+ takes a boolean value: if true, the program guesses the shapefile's projection. \verb+--shapefile.id-column+ is the name of the column containing the ID of each shape in the shapefile. The parameters \verb+-n+ and \verb+-o+ are respectively the road network definition and the output XML file that can be used with SUMO.

\begin{figure}[!ht]
    \centering
    \includegraphics[width=.85\columnwidth]{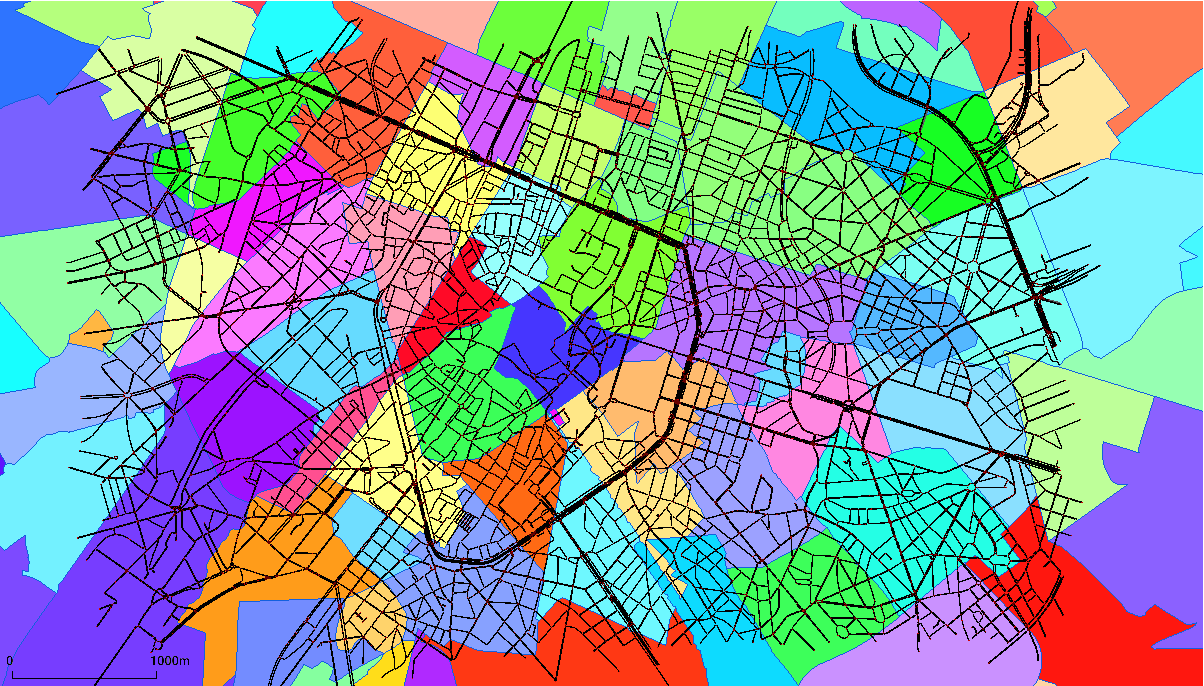}
    \caption{Spatial boundaries of Brussels capital region's neighborhoods in SUMO. Colors are assigned randomly to each neighborhood.}
    \label{fig:bxl_poly}
\end{figure}

Figure~\ref{fig:bxl_poly} shows the road network of Brussels city and the polygons delimiting the neighborhoods, obtained from the shapefile of CIRB. Each neighborhood is colored randomly. We used the \verb+netedit+ tool to visualize the road network and the neighborhood. 

SUMO provides the tool \verb+edgesInDistricts+\footnote{\href {https://sumo.dlr.de/docs/Tools/District.html}{https://sumo.dlr.de/docs/Tools/District.html}} to convert polygons delimiting the neighborhoods to TAZs. This tool reads the polygons from an input polygon files (created using \texttt{polyconvert}) and output an XML containing the TAZs definition. Each TAZ includes all the edges inside the polygon. The tool \verb+edgesInDistricts+ can be used as follow:
 
 \begin{lstlisting}
> python $\dollar$SUMO_HOME/tools/edgesInDistricts.py -n bxl.net.xml 
	-t taz.poly.xml -o TAZ.xml
 \end{lstlisting}
 
 The output file containing the TAZs has the following format:
 
 \begin{lstlisting}
<?xml version="1.0" encoding="UTF-8"?>
<tazs xmlns:xsi="http://www.w3.org/2001/XMLSchema-instance" 
  xsi:noNamespaceSchemaLocation="http://sumo.dlr.de/xsd/taz_file.xsd">
    <taz id="ANDERLECHT CENTRE - WAYEZ" color="51,128,255" edges="-1019451816#0
               -1019451816#1 -106463402#0 -106463402#1 ..."/>
    <taz id="ANNEESSENS" color="51,128,255" edges="-1007289475 ..."/>
    ... 
</tazs>
\end{lstlisting}

In case the polygons that separate the modeled urban area are not available, the TAZs can be generated randomly. SUMO provides two commands for this:

\begin{itemize}
	\item \verb+generateBidiDistricts.py+: create TAZs and assign to each one edges that are opposite of each other;
	\item \verb+gridDistricts.py+: create a grid of TAZs for an input road network, each one with a specified size (in meters).
\end{itemize}

\begin{figure}[!ht]
    \centering
    \includegraphics[width=.85\columnwidth]{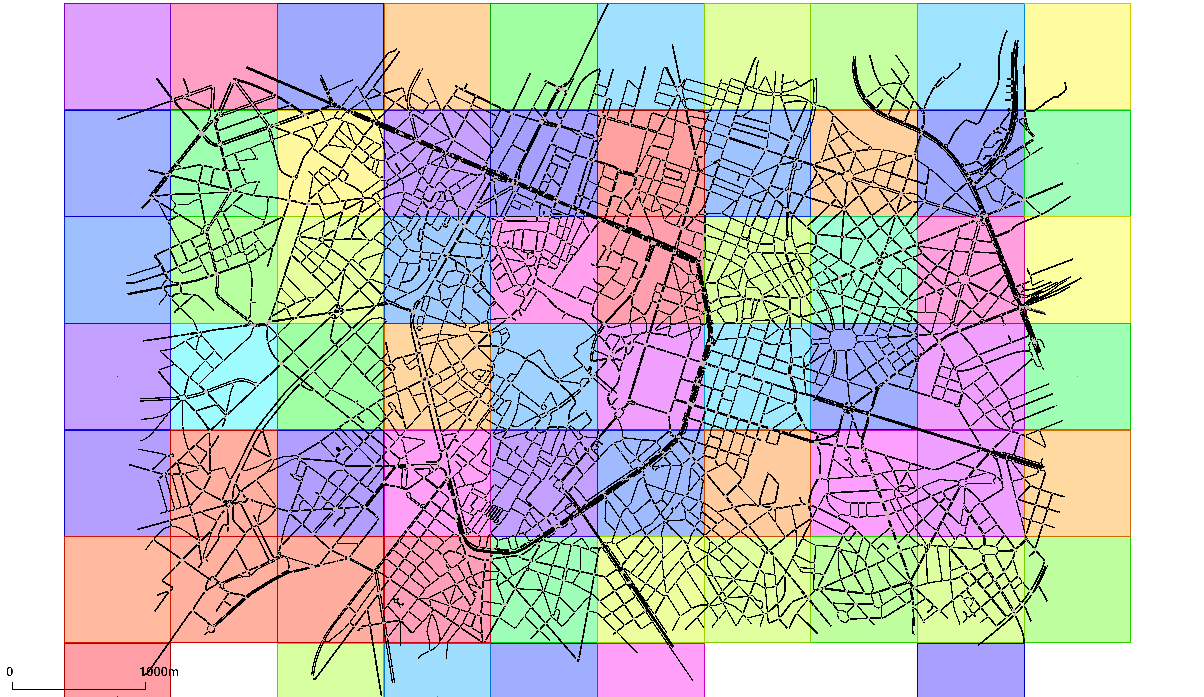}
    \caption{TAZs generated using the \texttt{gridDistricts.py} tool}
    \label{fig:bxl_taz_grid}
\end{figure}

Figure~\ref{fig:bxl_taz_grid} shows the TAZs generated randomly using the \verb+gridDistricts.py+ tool. 

\subsection{Trips Generation (step~\lgText{\threeCircled})}

Trips are the building blocks for modeling vehicular traffic. A trip is identified by a departure time, an origin and a destination point. A trip is associated with one unique vehicle. 

\subsubsection{Using OD-Based Traffic Demand}

Trips in SUMO can be modeled starting from Origin/Destination (OD) matrices. The content of an OD matrix is the number of vehicles that flow from an origin to a destination during a specific time horizon.


The command \verb+od2trips+\footnote{\href {https://sumo.dlr.de/docs/od2trips.html}{https://sumo.dlr.de/docs/od2trips.html}} imports OD matrices and splits them into individual vehicle trips. In the output file, each trip is defined by an id with starting and ending time (included inside the given time-lapse), and the origin and destination edges in the road network. If different transportation modes are considered, then \verb+od2trips+ must be used for each transportation mode and for each modeled time horizon. 


The following command produces a trip file for SUMO starting from an OD demand definition in file ``\texttt{OD\_matrix.od}'':

\begin{lstlisting}
> od2trips -v --taz-files TAZs.taz.xml --vtype passenger --prefix car 
	--od-matrix-files OD_matrix.od -o output/output.odtrips.xml
\end{lstlisting}

An OD matrix is typically in O-format, that lists each origin and each destination together with the amount of vehicles flowing from origin to destination. Following, we show an example of OD matrix in O-format.

\begin{lstlisting}
$\dollar$OR;D2
* From-Time  To-Time
7.00 8.00
* Factor
1.00
         1          1       1.00
         1          2       2.00
         1          3       3.00
         2          1       4.00
         2          2       5.00
         2          3       6.00
         3          1       7.00
         3          2       8.00
         3          3       9.00	
\end{lstlisting}

\noindent where:
\begin{itemize}
	\item The first line is a format specifier that must be included verbatim.
	\item The lines starting with `*' are comments and can be omitted
	\item The second non-comment line determines the time range given as \verb+HOUR.MINUTE HOUR.MINUTE+
	\item The third line is a global scaling factor for the number of vehicles for each cell
	\item All other lines describe matrix cells in the form \verb+FROM TO NUMVEHICLES+
\end{itemize}

\begin{figure}[!ht]
    \centering
    \includegraphics[width=0.7\linewidth]{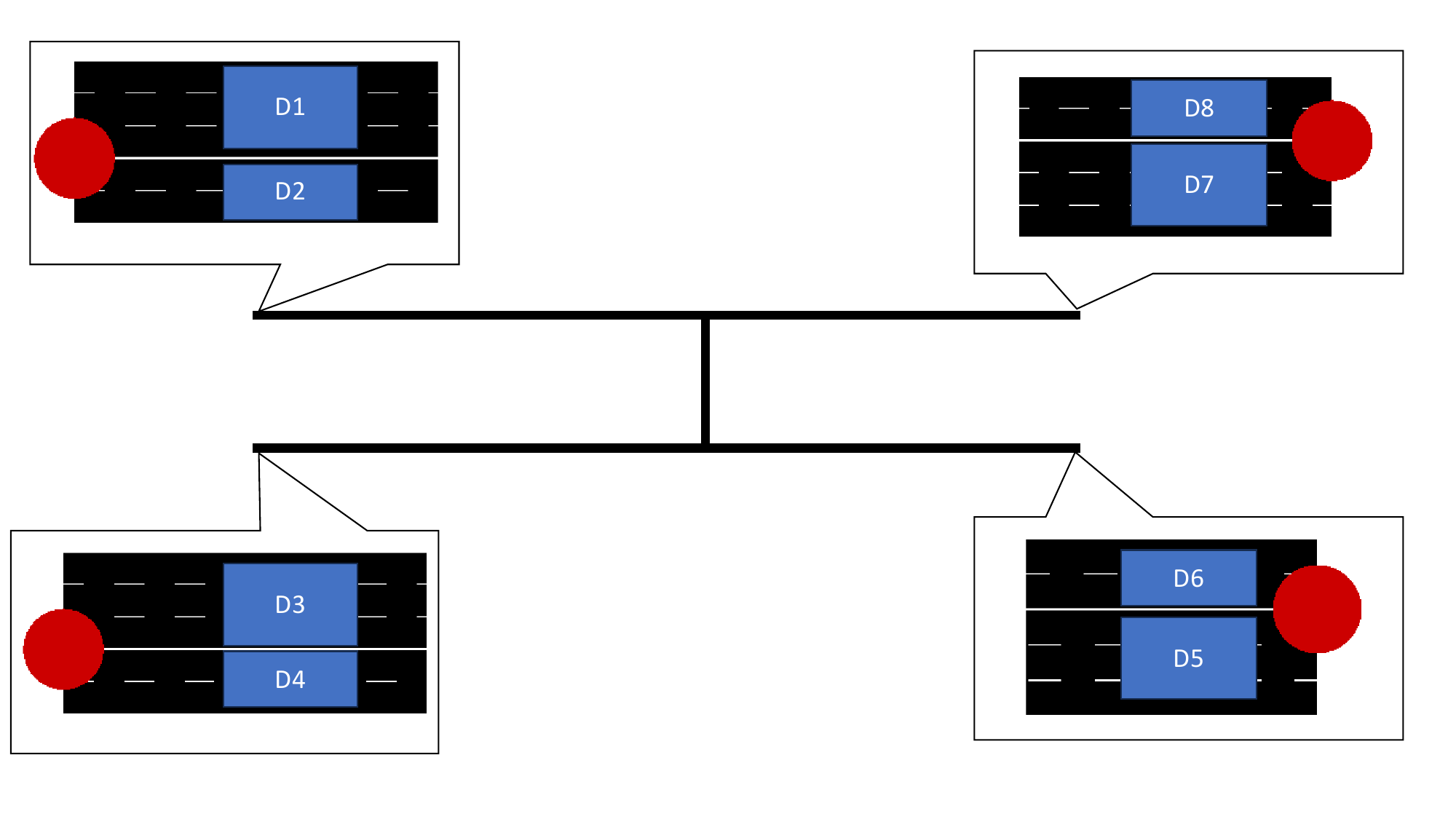}
    \caption{Road network example. Edges from D1 to D8 are used to model origin and destination of vehicles.}
    \label{fig:quickstart}
\end{figure}

A further TAZ format supported by SUMO is the \textbf{tazRelation}. The following example shows how to use OD traffic demand in the tazRelation format and generate a traffic demand compatible with SUMO. Figure~\ref{fig:quickstart} shows a sample road network. In this network, we consider eight edges as origins and destinations. 

Following, we report an example of OD demand definition in tazRelation format for the road network in Figure~\ref{fig:quickstart}.

\begin{verbatim}
<data>
    <interval id="0" begin="0" end="3600">
        <tazRelation from="D2" to="D7" count="100"/>
        <tazRelation from="D2" to="D5" count="30"/>
        <tazRelation from="D2" to="D3" count="90"/>
        <tazRelation from="D8" to="D1" count="90"/>
        <tazRelation from="D8" to="D3" count="200"/>
        <tazRelation from="D8" to="D5" count="50"/>
        <tazRelation from="D4" to="D1" count="10"/>
        <tazRelation from="D4" to="D7" count="170"/>
        <tazRelation from="D4" to="D5" count="90"/>
        <tazRelation from="D6" to="D3" count="110"/>
        <tazRelation from="D6" to="D7" count="30"/>
        <tazRelation from="D6" to="D1" count="80"/>
    </interval>
</data>
\end{verbatim}

The \texttt{from} and \texttt{to} fields indicate respectively the origin and destination edges. The field \texttt{count} is the number of vehicles that must flow from the origin to the destination edge during the interval ``0''. We now use the tool \texttt{od2trips} to imports OD demand in tazRelation and split them into individual vehicle trips. This tool generates an XML file that has the following format (using the previous TAZ definition):

\begin{lstlisting}
<routes>
    <trip id="63" depart="0.79" from="D2" to="D3" fromTaz="D2" toTaz="D3"/>
    <trip id="282" depart="2.83" from="D4" to="D5" fromTaz="D4" toTaz="D5"/>
    <trip id="516" depart="4.77" from="D6" to="D1" fromTaz="D6" toTaz="D1"/>
    <trip id="327" depart="8.55" from="D4" to="D7" fromTaz="D4" toTaz="D7"/>
    <trip id="244" depart="11.12" from="D4" to="D5" fromTaz="D4" toTaz="D5"/>
    <trip id="88" depart="16.72" from="D2" to="D3" fromTaz="D2" toTaz="D3"/>
    <trip id="9" depart="18.39" from="D2" to="D3" fromTaz="D2" toTaz="D3"/>
    ...
<routes/>
\end{lstlisting}

Differently from the tazRelation format, the file generated by \texttt{od2trips} contains information about each vehicle that should be inserted into the simulation. Next, we use the trip file in the subsequent traffic assignment (Section~\ref{sec:traffic_assignment}), followed by the simulation step. The traffic assignment involves finding a path from the origin to the destination edge.

SUMO also provides a tool that allows generating OD matrices from TAZs and route definition files, named \verb+route2OD.py+\footnote{\href {https://sumo.dlr.de/docs/Tools/Routes.html\#route2odpy}{https://sumo.dlr.de/docs/Tools/Routes.html\#route2odpy}}. Following we show an example of, requiring a route file (-r), and a TAZs file (-a).

\begin{lstlisting}
> python $\dollar$SUMO_HOME/tools/route/route2OD.py -r duarouter.rou.xml 
	-a TAZ.xml -o OD_matrix.xml
\end{lstlisting}

We provide a script to generate random OD traffic demand\footnote{\href {https://github.com/davide990/ODgenerator}{https://github.com/davide990/ODgenerator}} (called ODDemandGenerator.py), which can be used as follow:

\begin{lstlisting}
> python ODDemandGenerator.py -n my_net.net.xml -t taz.out.xml -p 10 
 -x odpairs.xml -l 0 -j 10
\end{lstlisting}

The output of this script are two: the TAZ definition file and the OD demand definition.

\subsubsection{Generating Random Trips}

SUMO provides the tool \verb+randomTrips.py+\footnote{\href {https://sumo.dlr.de/docs/Tools/Trip.html\#randomtripspy}{https://sumo.dlr.de/docs/Tools/Trip.html\#randomtripspy}} to generate random trips for a given road network. It does so by choosing source and destination edge either uniformly or at random. The resulting trips are stored in an XML file that can be later used with \verb+duarouter+\footnote{\href {https://sumo.dlr.de/docs/duarouter.html}{https://sumo.dlr.de/docs/duarouter.html}}, a tool available in SUMO for generating routes. The trips are distributed evenly in a temporal interval defined by begin (option -b, default 0) and end time (option -e, default 3600) in seconds. The number of trips is defined by the repetition rate (option -p, default 1) in seconds. 

The following command generates random trips starting in one hour interval for a given road network (parameter \verb+-n+):

\begin{lstlisting}
> python tools/randomTrips.py -n <net-file> -e 3600
\end{lstlisting}

The \verb+randomTrips+ command allows specifying the density of traffic generated per unit of time. By default, one vehicle is added into the simulation at each second. The period at which vehicles are added into the simulation can be specified by using the \verb+--period <FLOAT>+ option: in this way, the arrival rate is of (1/period) per second. By using values below 1, multiple vehicles are added into the simulation at each second.

If several \verb+<FLOAT>+ numbers are provided, like in \verb+--period 1.0 0.5+, the time interval will be divided equally into sub-intervals, and the arrival rate for each sub-interval is controlled by the corresponding period (in the preceding example, a period of 1.0 will be used for the first sub-interval and a period of 0.5 will be used for the second). There are two other ways to specify the insertion rate:

\begin{itemize}
	\item using the \verb+--insertion-rate+ parameter: this is the number of vehicles per hour that the user expects.
	\item using the \verb+--insertion-density+ parameter: this is the number of vehicles per hour per kilometer of road that the user expects (the total length of the road is computed with respect to a certain vehicle class that can be changed with the option --edge-permission).
\end{itemize}

When adding option \verb+--binomial <INT>+ the arrivals will be randomized using a binomial distribution where $n$ (the maximum number of simultaneous arrivals) is given by the argument to \verb+--binomial+ and the expected arrival rate is 1/period.

Let us suppose we want to let n vehicles depart between times \verb+t0+ and \verb+t1+ set the options, the following parameters must be provided:

\begin{lstlisting}
> python tools/randomTrips.py -n <net-file> -b t0 -e t1 -p ((t1 - t0) / n)
\end{lstlisting}

By default the departures of all vehicles are equally spaced in time. Since the inserted vehicle are spread randomly over the whole network, this comes out as a binomial distribution of inserted vehicles for each individual edge which gives a good approximation to the Poisson distribution if the network is large (and hence the insertion probability of each edge is small). By setting set option \verb+--random-depart+, the (still fixed) number of departure times are drawn from a uniform distribution over [begin, end]. This leads to an exponential distribution of insertion time headways between vehicles on all edges (which is the headway pattern of the Poisson distribution). Hence, this is useful to have a more varied insertion time pattern for small networks.

One interesting feature that is available in \verb+randomTrips+ is that it is possible to assign a unique weight to each edge into the input network by using the \verb+--weights-prefix <STRING>+ paraemter with the prefix value as argument.

The tool will load weights for all edges by finding a file (within the running directory) with extension .src.xml, .dst.xml or .via.xml. According to the file extension, weights are used differently for routes generation:

\begin{itemize}
	\item .src.xml contains the probabilities for an edge to be selected as from-edge
	\item .dst.xml contains the probabilities for an edge to be selected as to-edge
	\item .via.xml contains the probabilities for an edge to be selected as via-edge (only used when option \verb+--intermediate+ is set).
\end{itemize}

\subsection{Traffic Assignment (step~\lgText{\fourCircled})}\label{sec:traffic_assignment}

So far, we have described how to create trips, each specifying a vehicle's origin, destination, and the time at which it should be introduced into the traffic simulation. Now, we explain how to generate routes. A route consists of a sequence of edges that define the specific sections of the road network a vehicle must traverse during the simulation. In other words, a route must be assigned to each trip to ensure that vehicles can navigate from their origin to their destination.

Although in the scientific literature there are several vehicular routing algorithms~\cite{6398701}, herein we will use the routing method provided by SUMO through the \verb+duarouter+ tool. By default, this tool uses the Dijkstra to generate routes, but there are other methods available: A*, CH (Contraction Hierarchies), CH Wrapper.

The following command shows how to use the \verb|duarouter| tool to generate routes:

\begin{lstlisting}
> duarouter --net-file osm.net.xml --route-files od2trips.out.xml
    --output-file duarouter.rou.xml
\end{lstlisting}

\noindent where \verb|--net-file| is the file name of the road network, \verb|--route-files| it the file containing the trips.

The routing process can also be performed iteratively by using the \verb+duaIterate.py+\footnote{\href {https://sumo.dlr.de/docs/Tools/Assign.html\#duaiteratepy}{https://sumo.dlr.de/docs/Tools/Assign.html\#duaiteratepy}} script. This executes iteratively the following steps, in order:

\begin{enumerate}
	\item Execute \verb+duarouter+ to perform the (re-)routing 
	\item Calling SUMO to simulate travel times
\end{enumerate}

To optimize traffic flow, the \verb+duaIterate.py+ tool executes several simulations (the number of simulation can be configured). For each simulation, the tool finds a set of alternative routes that reduce the travel time for all vehicles. These routes are defined based on the traffic information produced by the simulation.

The following step consists of choosing routes in a set of alternatives. SUMO includes two route choice models: Gawron and Logit, both considering a weight/cost function.

Gawron proposes a probabilistic approach in which each route choice is modeled by a discrete probability distribution~\cite{gawron_iterative_1998}. The algorithm takes in input the following parameters:

\begin{itemize}
	\item the travel time along the used route in the previous simulation step,
	\item the sum of edge travel times for a set of alternative routes,
	\item the probability of choosing a route in the previous simulation step.
\end{itemize}

In each simulation, a driver $d$ chooses a route $r$ from a set $R_d$ (in which the driver is allowed to transit) according to the probability distribution $p_d$. Each route is associated with a cost $c_d : R_d \rightarrow \mathcal{R}_{+}$; this function estimates the time required by the driver to reach the destination. The driver cannot know in advance the amount of traffic in the road network; so, the idea is that the driver evaluate an estimate of the time needed, and updates incrementally (at each simulation) its knowledge about the time required to reach a certain destination.

In Logit method, the required amount of time to travel each road is calculated according to only the information from the last simulation: it ignores old costs and old probabilities and takes the route cost directly as the sum of the edge costs from the last simulation:

\begin{equation}
c'_r = \sum_{e \in r} w(e)
\end{equation}

\noindent where $c'_r$ is the updated cost of route $r$, $w(e)$ is the weight of the edge $e$ calculated from the input weight/cost function. The probability $p'_r$ for each route $r$ is calculated from an exponential function with parameter $\theta$ scaled by the sum over all route values:

\begin{equation}
p'_r = \frac{\textnormal{exp}\left(\theta c'_r\right)} {\sum_{s\in R} \textnormal{exp}\left(\theta c'_s\right)}
\end{equation}

\noindent where $c'_r$ is the cost of route $r$, $w(e)$ is the weight of the edge $e$ calculated from the input weight/cost function.

The following command shows how to use the \verb|duaIterate| tool to generate optimized routes:

\begin{lstlisting}
> python $\dollar$SUMO_HOME/tools/assign/duaIterate.py -n road_net.xml 
    -t trips.rou.xml -l 2
\end{lstlisting}

\noindent where \verb|-n| is the file name of the road network, \verb|-t| it the file containing the trips, \verb|-l| the number of iterations (using a number of iterations of 1 is equivalent to calling \verb|duarouter|). By default, the Gawron model is used as the default route choice model.

\subsection{Running the simulation (step~\lgText{\fiveCircled})}

\begin{figure}[!ht]
    \centering
    \includegraphics[width=.85\columnwidth]{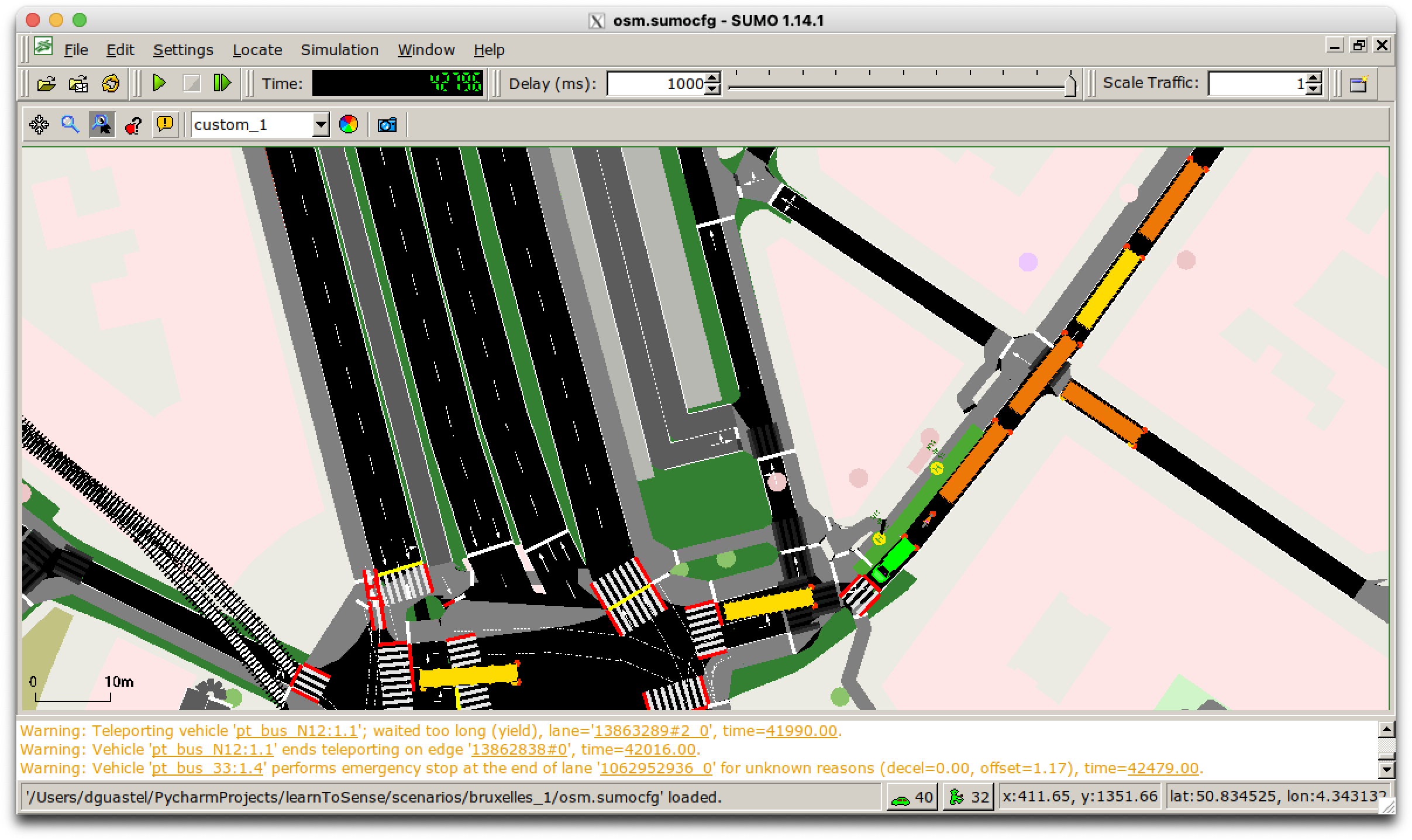}
    \caption{SUMO graphical user interface.}
    \label{fig:sumo_ui}
\end{figure}

SUMO provides two ways of executing simulations: from the command line (using the \verb+sumo+ command) and from the graphical interface (using the \verb+sumo-gui+ command). Both commands use the same set of parameters, including the road network, routes, and the time horizon over which the simulation is conducted. Alternatively, a unique configuration file can be provided as parameter (using the parameter\texttt{-c}), containing all the required parameters and files for running a simulation. Following, we show an example of a configuration file:

\begin{lstlisting}
<configuration>
    <input>
        <net-file value="test.net.xml"/>
        <route-files value="test.rou.xml"/>
        <additional-files value="test.add.xml"/>
    </input>
</configuration>
\end{lstlisting}

\noindent SUMO can be launched using the following command (assuming that the previous configuration file is named test.sumocfg):

\begin{lstlisting}
> sumo -c test.sumocfg
\end{lstlisting}

Optionally, SUMO provides a graphical interface, available using the \verb+sumo-gui+ command (instead of \verb|sumo|). The interface provides a graphical way to monitoring traffic during the simulation. Figure~\ref{fig:sumo_ui} shows the SUMO graphical user interface.

\section{Output Configuration (step~\lgText{\sixCircled})}\label{sec:output} 

SUMO allows generating output files containing several measures about traffic. These output files can be used to analyze traffic patterns or congestion phenomena.


A complete list of the output files that can be produced by the simulator is available at the SUMO website\footnote{\url{https://sumo.dlr.de/docs/Simulation/Output/index.html}}.

\subsection{Virtual Traffic Monitoring Sensors}

SUMO allows adding virtual monitoring devices that produce traffic counts. This is to simulate the behavior of traffic monitoring devices such as camera or induction loops. One type of virtual sensor, known in SUMO as \textbf{laneAreaDetector}, monitor traffic along one specific lane. Their functioning is similar to traffic cameras.

\begin{figure}[!ht]
    \centering
    \includegraphics[width=.95\columnwidth]{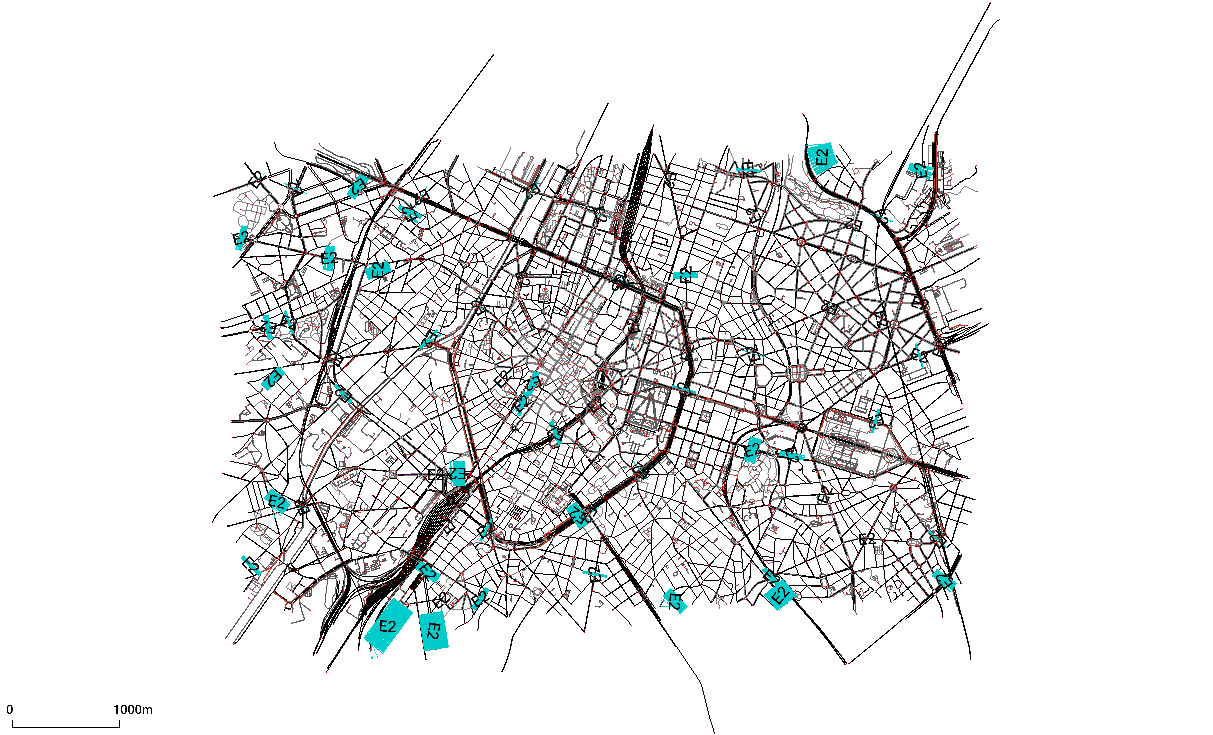}
    \caption{Position of lane area detectors (virtual sensors), colored in green, in a model of the city of Brussels.}
    \label{fig:sumo_lanedetectors}
\end{figure}

A laneAreaDetector can be defined using the \verb+netedit+\footnote{\href {https://sumo.dlr.de/docs/Netedit/index.html}{https://sumo.dlr.de/docs/Netedit/index.html}} tool, or using a specific XML definition. Figure~\ref{fig:sumo_lanedetectors} shows a map of Brussels city where multiple laneAreaDetector sensors (in teal color) have been placed arbitrarily.

After the definition of virtual sensors in \verb+netedit+, the set of sensors can be exported as an XML file, which has the following format:

\begin{lstlisting}
<?xml version="1.0" encoding="UTF-8"?>
<additional xmlns:xsi="http://www.w3.org/2001/XMLSchema-instance" 
	xsi:noNamespaceSchemaLocation="http://sumo.dlr.de/xsd/additional_file.xsd">
    <laneAreaDetector id="e2_0" lane="-159327011#2_0" pos="0.0" 
    		length="138.630000" period="100" file="detectors.out.xml"/>
    <laneAreaDetector id="e2_1" lane="15497309#1_0" pos="0.0" 
    		length="1.940000" period="100" file="detectors.out.xml"/>
	...
</additional>
\end{lstlisting}

\noindent where the parameter \texttt{file} is the path to the file that contains the traffic counts produced by virtual sensors. To use virtual sensors in SUMO, the XML file containing their definition must be provided as input to the simulator. At the end of a simulation, the file indicated in field \texttt{file} (in the previous example, ``detectors.out.xml'') contains the traffic counts observed by each sensor during the indicated time period. Table~\ref{tab:laneareadetectors_types} reports the list of information generated by a virtual sensor. 

\begin{table}[!ht]
\centering
\caption{Subset of data types returned by each laneAreaDetector at the end of a simulation.}
\label{tab:laneareadetectors_types}
\begin{tabular}{|l|l|l|}
\hline
\textbf{Name}    & \textbf{Unit}                                                   & \textbf{Description}                                                                                                                                                                              \\ \hline
begin            & \begin{tabular}[c]{@{}l@{}}(simulation) \\ seconds\end{tabular} & The first time step the values were collected in                                                                                                                                                  \\ \hline
end              & \begin{tabular}[c]{@{}l@{}}(simulation) \\ seconds\end{tabular} & \begin{tabular}[c]{@{}l@{}}The last time step + DELTA\_T the values were \\ collected in (may be equal to begin)\end{tabular}                                                                     \\ \hline
id               & id                                                              & \begin{tabular}[c]{@{}l@{}}The id of the detector (needed if several detectors \\ share an output file)\end{tabular}                                                                              \\ \hline
meanSpeed        & m/s                                                             & The mean velocity over all collected data samples.                                                                                                                                                \\ \hline
meanTimeLoss     & s                                                               & \begin{tabular}[c]{@{}l@{}}The average time loss per vehicle in the corresponding \\ interval. The total time loss can be obtained by \\ multiplying this value with nVehSeen.\end{tabular}       \\ \hline
meanOccupancy    & \%                                                              & \begin{tabular}[c]{@{}l@{}}The percentage (0-100\%) of the detector's place that was \\ occupied by vehicles, summed up for each time step and \\ averaged by the interval duration.\end{tabular} \\ \hline
maxOccupancy     & \%                                                              & \begin{tabular}[c]{@{}l@{}}The maximum percentage (0-100\%) of the detector's place \\ that was occupied by vehicles during the interval.\end{tabular}                                            \\ \hline
maxVehicleNumber & \#                                                              & \begin{tabular}[c]{@{}l@{}}The maximum number of vehicles that were on the detector\\  area during the interval.\end{tabular}                                                                     \\ \hline
\end{tabular}
\end{table}

To use the virtual sensors in a simulation, the path to the XML file containing the definition of laneAreaDetectors must be specified inside the \verb|<additional>| tag of a SUMO configuration file (*.sumocfg), or passed as parameter (\verb|-a|).

We developed a tool in python language that enables generating random laneAreaDetectors in a given road network\footnote{\href {https://gitlab.com/traffic-sim/random-lane-detector-placer}{https://gitlab.com/traffic-sim/random-lane-detector-placer}}. The tool performs the following steps, in order:

\begin{enumerate}
	\item extract the TAZs for the modeled road network (see Section~\ref{sec:taz});
	\item extract all the edges and assign them to TAZs;
	\item For each TAZ:
	\begin{enumerate} 
	\item choose one random location where to put a laneAreaDetector with probability $p$;
	\item Place the laneAreaDetectors into the network.
	\end{enumerate}
\end{enumerate}

The probability $p$ is calculated according to two strategies:

\begin{itemize}
    \item \textbf{by number of lanes}: an edge has a probability of getting a virtual sensor proportional to the number of lanes. 
    \item \textbf{by weight}: the probability of an edge of getting a virtual sensor depends on a parameter specified manually, and which values are in the `edgedata' output file produced by SUMO\footnote{\href {https://sumo.dlr.de/docs/Simulation/Output/Lane-\_or\_Edge-based\_Traffic\_Measures.html}{https://sumo.dlr.de/docs/Simulation/Output/Lane-\_or\_Edge-based\_Traffic\_Measures.html}}.
\end{itemize}








\section{Tools for the Automatic Definition of Scenarios}\label{sec:tools} 

\subsection{SUMO OSM Web Wizard}

\begin{figure}[!ht]
    \centering
    \includegraphics[width=.9\columnwidth]{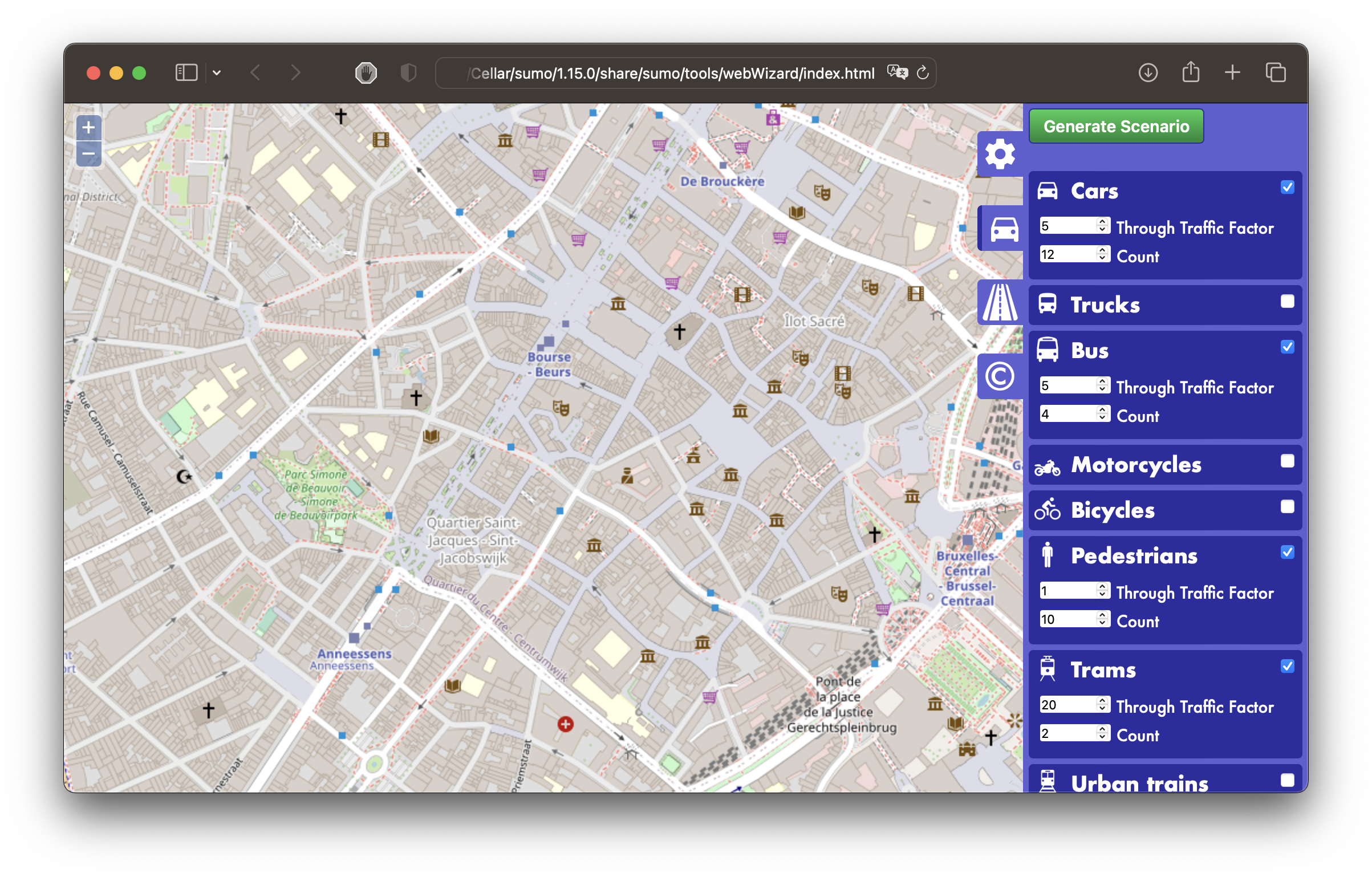}
    \caption{SUMO OSM Web Wizard user interface.}
    \label{fig:osm_wizard}
\end{figure}

The OSM Web Wizard\footnote{\href {https://sumo.dlr.de/docs/Tutorials/OSMWebWizard.html}{https://sumo.dlr.de/docs/Tutorials/OSMWebWizard.html}} is a web-based tool that offers an easy solutions to start modeling traffic scenarios with SUMO. The user can specify the area to model graphically through an openstreetmap map excerpt, and configure a randomized traffic demand and run and visualize the scenario in the sumo-gui. To run this tool, the following command must be run:

\begin{lstlisting}
> python $\dollar$SUMO_HOME/tools/osmWebWizard.py
\end{lstlisting}

Figure~\ref{fig:osm_wizard} shows the OSM Web Wizard interface.

\subsection{SAGA}

SUMO Activity GenerAtion (SAGA) is a user-defined activity-based multimodal mobility scenario generator for SUMO~\cite{codeca_saga_2022}. SAGA is capable of handling activity-based mobility from detailed information on the environment (e.g., buildings, PoIs), as well as the transportation infrastructure. SAGA is capable of extracting environmental information available automatically (from OSM) such as building and public transportation lines, generating all the configuration files required by SUMO, and fills the missing information with sensible default values. The output scenario is multi-modal, that is, including different types of transportation mean for population, and includes also parking areas, buildings, Points of Interest (PoIs).

\bibliographystyle{plain} 
\bibliography{main}

\begin{appendices}

\section{Full Step-by-Step Examples}

\subsection{Example 1}

\noindent\textbf{\framebox{Step 1} -- Network generation}

\begin{lstlisting}
netgenerate --rand -o MySUMOFile.net.xml --rand.iterations=200 
	-j traffic_light --random
\end{lstlisting}

\noindent\textbf{\framebox{Step 2} -- Random trips generation}

\begin{lstlisting}
python $\dollar$SUMO_HOME/tools/randomTrips.py -n MySUMOFile.net.xml -e 3600 
	-o trips.rou.xml --random
\end{lstlisting}

\noindent\textbf{\framebox{Step 3} -- Routing}

\begin{lstlisting}
duarouter --net-file MySUMOFile.net.xml --route-files trips.rou.xml 
	--output-file duarouter.rou.xml --ignore-errors
\end{lstlisting}

\noindent\textbf{\framebox{Step 4} -- Configuration file generation}

\begin{lstlisting}
echo "<configuration>
    <input>
        <net-file value=\"MySUMOFile.net.xml\"/>
        <route-files value=\"duarouter.rou.xml\"/>
    </input>
</configuration>" $\mbox{|}$ tee -a sumo.sumocfg
\end{lstlisting}

\noindent\textbf{\framebox{Step 5} -- Simulation}

\begin{lstlisting}
sumo-gui -c sumo.sumocfg
\end{lstlisting}

\subsection{Example 2}

\noindent\textbf{\framebox{Step 1} -- Network generation}

\begin{lstlisting}
netgenerate --rand -o MySUMOFile.net.xml --rand.iterations=300 
	-j traffic_light --random --rand.grid
\end{lstlisting}

\noindent\textbf{\framebox{Step 2} -- Random trips generation}

\begin{lstlisting}
python $\dollar$SUMO_HOME/tools/randomTrips.py -n MySUMOFile.net.xml -e 3600 
	-o trips.rou.xml --random --random-depart --binomial 20
\end{lstlisting}

\noindent\textbf{\framebox{Step 3} -- Routing}

\begin{lstlisting}
duarouter --net-file MySUMOFile.net.xml --route-files trips.rou.xml 
	--output-file duarouter.rou.xml --ignore-errors --route-choice-method logit
\end{lstlisting}

\noindent\textbf{\framebox{Step 4} -- Configuration file generation}

\begin{lstlisting}
echo "<configuration>
    <input>
        <net-file value=\"MySUMOFile.net.xml\"/>
        <route-files value=\"duarouter.rou.xml\"/>
    </input>
</configuration>" | tee -a sumo2.sumocfg	
\end{lstlisting}

\noindent\textbf{\framebox{Step 5} -- Simulation}

\begin{lstlisting}
sumo-gui -c sumo2.sumocfg
\end{lstlisting}

\subsection{Example 3}

\noindent\textbf{\framebox{Step 1} -- Get the map from OSM}

\begin{lstlisting}
python $\dollar$SUMO_HOME/tools/osmGet.py --bbox="4.3583, 50.8362, 4.3696, 50.8455"
\end{lstlisting}

\noindent\textbf{\framebox{Step 2} -- OSM map to SUMO network}

\begin{lstlisting}
python $\dollar$SUMO_HOME/tools/osmBuild.py --osm-file osm_bbox.osm.xml
\end{lstlisting}

\noindent\textbf{\framebox{Step 3} -- Extract the polygons from the OSM file}

\begin{lstlisting}
polyconvert --net-file osm.net.xml --osm-files osm_bbox.osm.xml 
	--type-file $\dollar$SUMO_HOME/share/sumo/data/typemap/osmPolyconvert.typ.xml 
	-o polygons.poly.xml
\end{lstlisting}

\noindent\textbf{\framebox{Step 4} -- Generate random trips}

\begin{lstlisting}
python $\dollar$SUMO_HOME/tools/randomTrips.py -n osm.net.xml -e 3600 -o trips.rou.xml 
	--random --random-depart --binomial 10 --validate
\end{lstlisting}

\noindent\textbf{\framebox{Step 5} -- Routing}

\begin{lstlisting}
duarouter --net-file osm.net.xml --route-files trips.rou.xml 
	--output-file duarouter.rou.xml --ignore-errors true
\end{lstlisting}

\noindent\textbf{\framebox{Step 6} -- Configuration file generation}
\begin{lstlisting}
echo "<configuration>
    <input>
        <net-file value=\"osm.net.xml\"/>
        <route-files value=\"duarouter.rou.xml\"/>
        <additional-files value=\"polygons.poly.xml\"/>
    </input>
</configuration>" | tee -a sumo.sumocfg
\end{lstlisting}

\noindent\textbf{\framebox{Step 7} -- Simulation}

\begin{lstlisting}
sumo-gui -c sumo2.sumocfg
\end{lstlisting}

\subsection{Example 4}

\noindent\textbf{\framebox{Step 1} -- Scenario Definition}

\begin{lstlisting}
python $\dollar$SUMO_HOME/tools/osmGet.py --bbox="4.3583, 50.8362, 4.3696, 50.8455"
python $\dollar$SUMO_HOME/tools/osmBuild.py --osm-file osm_bbox.osm.xml
\end{lstlisting}
\bigskip

\noindent\textbf{\framebox{Step 2} -- TAZs Extraction}
\begin{lstlisting}
polyconvert --net-file osm.net.xml --osm-files osm_bbox.osm.xml 
    --type-file $\dollar$SUMO_HOME/share/sumo/data/typemap/osmPolyconvert.typ.xml 
    -o polygons.taz.xml --type taz
python $\dollar$SUMO_HOME/tools/edgesInDistricts.py 
    -n osm.net.xml -t polygons.taz.xml -o TAZ.xml 
    -l passenger --complete
\end{lstlisting}
\bigskip

\noindent\textbf{\framebox{Step 3} -- Random Traffic Generation}
\begin{lstlisting}
python $\dollar$SUMO_HOME/tools/randomTrips.py -n osm.net.xml 
    -b 0 -e 3600 -o trips.rou.xml --random --random-depart 
    --binomial 10 --validate
\end{lstlisting}
\bigskip

\noindent\textbf{\framebox{Step 4} -- Random traffic to OD-matrix}
\begin{lstlisting}
python $\dollar$SUMO_HOME/tools/route/route2OD.py 
    -r trips.rou.xml -a TAZ.xml -o routes.xml -i 5
\end{lstlisting}
\bigskip

\noindent\textbf{\framebox{Step 5} -- Importing OD-matrix}
\begin{lstlisting}
od2trips -z routes.xml -n TAZ.xml -b 0 -e 3600 
    -o od2trips.out.xml
\end{lstlisting}
\bigskip

\noindent\textbf{\framebox{Step 6} -- Routing and Simulation}
\begin{lstlisting}
duarouter --net-file osm.net.xml --route-files od2trips.out.xml 
    --output-file duarouter.rou.xml --ignore-errors true
sumo-gui +a TAZ.xml -n osm.net.xml -r od2trips.out.xml -b 0 
    -e 3600 --edgedata-output edgedata.outsampler.xml --ignore-route-errors
\end{lstlisting}

\subsection{Example 5}

This example shows how to generate an inter-modal traffic scenario. We assume that the file ``map.osm'' is the map downloaded from OSM.
\bigskip

\noindent\textbf{\framebox{Step 1} -- OSM to SUMO, Public Transportation Data Extraction}
\begin{lstlisting}
netconvert --osm-files map.osm -o bxl.net.xml --osm.stop-output.length 20 
    --ptstop-output additional.xml --ptline-output ptlines.xml
\end{lstlisting}
\bigskip

\noindent\textbf{\framebox{Step 2} -- Find Travel Times and Create Public Transportation Schedules}

\begin{lstlisting}
python $\dollar$SUMO_HOME/tools/ptlines2flows.py -n bxl.net.xml -s additional.xml 
    -l ptlines.xml -o flows.rou.xml -p 600 --use-osm-routes --ignore-errors
\end{lstlisting}
\bigskip

\noindent\textbf{\framebox{Step 3} -- Generate Random Traffic for Vehicles}

\begin{lstlisting}
python $\dollar$SUMO_HOME/tools/randomTrips.py -n bxl.net.xml -b 0 -e 3600 
    -o passenger_trips.rou.xml --random --random-depart --validate 
    --vehicle-class passenger
\end{lstlisting}
\bigskip

\noindent\textbf{\framebox{Step 4} -- Vehicles Routing}

\begin{lstlisting}
duarouter --net-file bxl.net.xml --route-files passenger_trips.rou.xml 
    --output-file duarouter_passenger.rou.xml --ignore-errors
\end{lstlisting}
\bigskip

\noindent\textbf{\framebox{Step 5} -- Simulation}

\begin{lstlisting}
sumo-gui -n bxl.net.xml -r duarouter_passenger.rou.xml,flows.rou.xml -b 0 
    -e 3600 --edgedata-output edgedata.outsampler.xml 
    -a additional.xml --ignore-route-errors
\end{lstlisting}
\bigskip

\section{Reverting Road Direction in OSM} \label{app:reverse_road}

The following python method can be used to invert the direction of the edges whose ID is given in input. This method works only if the SUMO road network has been converted from the OSM using the \texttt{--output.original-names}.

\begin{lstlisting}[language=Python]
    def reverse_roads(edge_ids:list[String], osm_path:String) -> None:
    tree = ET.parse(osm_path)
    root = tree.getroot()
    for child in root:
        if "id" not in child.attrib:
            continue
        for edge_id in edge_ids:
            if re.search(child.attrib["id"], edge_id):
                found_oneway_tag = False
                for tag_d in list(child.iter('tag')):
                    if tag_d.attrib["k"] == 'oneway':
                        found_oneway_tag = True
                        if tag_d.attrib['v'] == '-1':
                            tag_d.attrib['v'] = 'yes'
                        else:
                            tag_d.attrib['v'] = '-1'
                if not found_oneway_tag:
                    # maybe it's a 2direction street
                    removed = False
                    for tag_d in list(child.iter('tag')):
                        if tag_d.attrib["k"] == 'lanes' or tag_d.attrib["k"] == 'oneway':
                            child.remove(tag_d)
                            removed = True
                    if not removed:
                        item = ET.SubElement(child, "tag")
                        item.attrib["k"] = "lanes"
                        item.attrib["v"] = str(len(re.findall(child.attrib["id"], edge_id)))
                        item = ET.SubElement(child, "tag")
                        item.attrib["k"] = "oneway"
                        item.attrib["v"] = "-1"

    with open("{}.rev.osm".format(osm_path), 'wb') as f:
        tree.write(f, encoding='utf-8')
\end{lstlisting}

The method takes in input a list of SUMO edges ID, and the full path to the OSM map. A for loop iterates over all the OSM nodes having an ID. Then, we check if the ID of the current node is present in the input list \texttt{edge\_ids}. If so, The node corresponds to an edge whose direction must be inverted. If the road is one way, we change the value from -1 to 'yes' (or the opposite) to invert the direction. If the road has more than 2 lanes, then we search for the XML tag 'oneway', and if it's found, it is removed from the XML tree. Otherwise, we change the value of 'oneway' attribute as before.

\section{SUMO Installation}

Detailed installation instructions are available in the SUMO website: \url{https://sumo.dlr.de/docs/Installing/index.html}

For OSX, we suggest using Homebrew (\url{https://brew.sh/}) to install SUMO. We recommend to use the following commands, in order:

\begin{lstlisting}
>	brew install --cask xquartz
>	brew tap dlr-ts/sumo
>	brew install --with-examples --with-ffmpeg 
		--with-gdal --with-gl2ps --with-swig sumo
\end{lstlisting}

After the installation, the system variable \verb+SUMO_HOME+ must be set (system-wide) to the path containing SUMO. Typically the path is as the following: \url{/opt/homebrew/Cellar/sumo/1.XX.XX/}.

\section{Common Issues}

\subsection{Broken Folder Links When Using SUMO on OSX}

When using SUMO installed through homebrew into the path \url{/opt/homebrew/Cellar/sumo/1.XX.XX/}, the following commands must be run to fix broken links in SUMO folders:

\begin{lstlisting}
	sudo ln -s $\dollar$SUMO_HOME/share/sumo/tools/ $\dollar$SUMO_HOME/
	sudo ln -s $\dollar$SUMO_HOME/share/sumo/data $\dollar$SUMO_HOME/
\end{lstlisting}



\end{appendices}

\section{Useful Resources}

\begin{itemize}
	\item \url{https://www.eltis.org/sites/default/files/tool/conduits_key_performance_indicators_its.pdf}
	\item \url{https://www.arcgis.com/home/item.html?id=f40c3fdb26c74e64af4f2cb8311c5559}
	\item \url{https://monitoringdesquartiers.brussels/maps/statistiques-population-bruxelles/evolution-population/densite-de-population/1/2019/#}
	\item \url{https://www.acea.auto/files/ACEA-report-vehicles-in-use-europe-2022.pdf}
	\item Number of households with cars (Belgium): \url{https://statbel.fgov.be/en/themes/datalab/vehicles-household}
\end{itemize}








\end{document}